\documentclass[fleqn,usenatbib]{mnras}

\usepackage{newtxtext,newtxmath}

\usepackage[T1]{fontenc}

\DeclareRobustCommand{\VAN}[3]{#2}
\let\VANthebibliography\thebibliography
\def\thebibliography{\DeclareRobustCommand{\VAN}[3]{##3}\VANthebibliography}

\usepackage{graphicx}	
\usepackage{amsmath}	

\title[RSG Shock Cooling Model: I]{Shock cooling emission from explosions of red super-giants: I. A numerically calibrated analytic model}

\author[Morag et al.]{
Jonathan Morag\thanks{E-mail: jonathan.morag@weizmann.ac.il},$^{1}$
Nir Sapir,$^{1,2}$
Eli Waxman$^{1}$
\\
$^{1}$Weizmann Institute of Science, Rehovot, Israel\\
$^{2}$Soreq Nuclear Center, Nahal Soreq, Israel\\
}

\date{Accepted 2023 March 18. Received 2023 March 8; in original form 2022 August 22}

\pubyear{2022}

\begin{document}
\label{firstpage}
\pagerange{\pageref{firstpage}--\pageref{lastpage}}
\maketitle

\begin{abstract}
Supernova light curves are dominated at early time, hours to days, by photons escaping from the expanding shock heated envelope. We provide a simple analytic description of the time dependent luminosity, $L$, and color temperature, $T_{\rm col}$, for explosions of red supergiants (with convective polytropic envelopes), valid up to H recombination ($T\approx0.7$~eV). The analytic description interpolates between existing expressions valid at different (planar then spherical) stages of the expansion, and is calibrated against numerical hydrodynamic diffusion calculations for a wide range of progenitor parameters (mass, radius, core/envelope mass and radius ratios, metalicity), and explosion energies. The numerically derived $L$ and $T_{\rm col}$ are described by the analytic expressions with $10\%$ and $5\%$ accuracy respectively.
$T_{\rm col}$ is inferred from the hydrodynamic profiles using frequency independent opacity, based on tables we constructed for this purpose (and will be made publicly available) including bound-bound and bound-free contributions. In an accompanying paper (Paper II) we show, using a large set of multi-group photon diffusion calculations, that the spectral energy distribution is well described by a Planck spectrum with $T=T_{\rm col}$, except at UV frequencies, where the flux can be significantly suppressed due to strong line absorption. We defer the full discussion of the multi-group results to paper II, but provide here for completeness an analytic description also of the UV suppression. Our analytic results are a useful tool for inferring progenitor properties, explosion velocity, and also relative extinction based on early multi-band shock cooling observations of supernovae.

\end{abstract}

\begin{keywords}
radiation: dynamics – shock waves – supernovae: general
\end{keywords}



\section{Introduction}

In a supernova explosion, the envelope of the progenitor star is expelled by an outward propagating radiation mediated shock (RMS). When the optical depth of the plasma lying ahead of the shock drops to $c/v_s$, where $v_s$ is the (instantaneous) shock velocity, photons contained in the shock transition layer escape and the RMS dissolves. In the absence of an optically thick circum-stellar medium this breakout takes place when the shock approaches the edge of the star, producing an X-ray/UV pulse of duration comparable to the larger of the light crossing time of the stellar radius, $R/c$, and the intrinsic pulse duration (given by the shock crossing time of the transition layer). The relatively short breakout pulse is followed by UV/optical "shock cooling" emission on a time scale of days, during which photons escape from deeper layers in the envelope as it expands \citep[see][for reviews]{waxman_shock_2016,levinson_physics_2020}. Observations of the early shock breakout and shock cooling emission may enable the determination of progenitor parameters, including its radius and envelope composition and mass, and of the ejecta velocity. The advantages of using such early observations are that they enable a direct determination of $R$, and that the predicted emission may be accurately calculated for given model parameters, thus enabling to accurately infer the parameter values from observations. The latter advantage is due mainly to the fact that the plasma is highly ionized and near local thermal equilibrium (LTE), which reduces opacity uncertainties and also enables the derivation of accurate analytic approximations. The challenge for using this method is obtaining high cadence, $\sim1$~hr, multi-band observations.

Due to the short duration of the breakout pulse, only a handful of cases have been observed with an indication for a breakout signal  \citep[e.g.][]{campana_association_2006,soderberg_extremely_2008,schawinski_supernova_2008,gezari_probing_2008,gezari_galex_2010,gezari_galex_2015}. Independent of shock breakout detections, observations of the shock cooling emission enable one to determine the progenitor parameters (including radius, envelope mass and composition) and ejecta velocity \citep{waxman_shock_2016,levinson_physics_2020}, and also the relative extinction \citep{rabinak_early_2011}. An accurate determination of these parameters requires early high cadence multi-band observations, including in particular at short, UV, wavelengths. UV measurements are essential for an accurate determination of the high color temperature, $T_{\rm col}$ \citep{rabinak_early_2011,sapir_uv/optical_2017,sagiv_science_2014,rubin_exploring_2017}. High cadence multi-band observations are required for a determination of the relative extinction \citep{rabinak_early_2011}, which strongly affects the inferred $T_{\rm col}$ and $L_{\rm bol}$. Most current observations do not satisfy these stringent requirements, and therefore typically enable only rough estimates of the parameters
\citep[e.g.][]{arcavi_sn_2011,sapir_uv/optical_2017,yaron_confined_2017,hosseinzadeh_short-lived_2018,andrews_sn_2019,soumagnac_sn_2020,dong_supernova_2021,tartaglia_early_2021,hosseinzadeh_weak_2022,gill_constraining_2022,jacobson-galan_circumstellar_2022,gagliano_early-time_2022}. It should be pointed out that ignoring extinction, or applying the model at times which are outside the range of its validity (e.g. due to low envelope masses), may lead to large systematic errors when inferring progenitor and explosion parameters from observations.

As the capabilities of rapid transient searches improve, e.g. with the Zwicky Transient Factory (ZTF) \citep{gal-yam_real-time_2011}, the upcoming Vera Rubin Observatory \citep{ivezic_lsst_2019}, and the expected launch of the wide-field UV space telescope ULTRASAT \citep{sagiv_science_2014}, the quantity and quality of data are expected to improve significantly. This will enable a systematic accurate determination of progenitor and explosion parameters based on shock cooling, and possibly shock breakout, observations. An accurate analytic description of the predicted flux is essential for this purpose. While detailed numeric calculations may provide higher accuracy predictions for specific model parameters, an analytic description is highly valuable for the "inverse problem" of inferring model parameters from observations: It does not require resource consuming numeric calculations for the derivation of model predictions for each point in the parameter space, and it enables a direct determination of degeneracies between and uncertainties of inferred model parameter values.

The indication that pre-SN precursors are common for IIn SNe on a month time scale preceding the explosion \citep[][and references therein]{ofek_precursors_2014}, provides evidence for intense mass loss episodes in many SN progenitors shortly before the explosion. In cases where the optical depth of the circum-stellar material (CSM) ejected from the progenitor star is larger than $c/v_s$, shock breakout may take place at large radii (compared to $R$), and the breakout time scale may be extended to days \citep[][and references therein]{waxman_shock_2016}. The current paper focuses on explosions of progenitors for which the CSM column density and optical depth are low, and the emission of radiation is dominated by the expanding stellar envelope.

\subsection{Earlier work}

\defcitealias{sapir_non-relativistic_2011}{SKW I}
\defcitealias{katz_non-relativistic_2012}{KSW II}
\defcitealias{sapir_non-relativistic_2013}{SKW III}
\defcitealias{rabinak_early_2011}{RW11}
\defcitealias{sapir_uv/optical_2017}{SW17}
\defcitealias{nakar_early_2010}{NS10}
\defcitealias{shussman_type_2016}{SWN16}
\defcitealias{piro_shock_2021}{PHY21}
\defcitealias{morag_shock_2022-1}{Paper II}
Exact numeric solutions and approximate analytic solutions for the shock breakout emission and for the subsequent shock cooling emission during the "planar" phase, at which the thickness of the emitting shell is small compared to the stellar radius, were derived by \citet{sapir_non-relativistic_2011}, \citet{katz_non-relativistic_2012} and  \citet{sapir_non-relativistic_2013} (hereafter \citetalias{sapir_non-relativistic_2011}, \citetalias{katz_non-relativistic_2012} and \citetalias{sapir_non-relativistic_2013} respectively). Approximate analytic solutions for the spherical phase, during which the emitting shell expands to radii much larger than the stellar radius, were derived in \citet{nakar_early_2010,rabinak_early_2011,nakar_supernovae_2014,piro_using_2015,sapir_uv/optical_2017,shussman_type_2016,piro_shock_2021}. With the exception of the work of \citet{piro_using_2015}, analytic and semi-analytic results were derived for polytropic envelopes \citep[][hereafter \citetalias{piro_shock_2021}, describe the velocity dependence of the density of the expanding envelope as a broken power-law, which approximately corresponds to the outcome of shock breakout from polytropic envelopes]{piro_shock_2021}. The predicted $L$ and $T_{\rm col}$ were found to be nearly independent of the polytropic index, and \citetalias{sapir_uv/optical_2017} have demonstrated that the properties of the post-breakout cooling emission are only weakly dependent on the density profile. The analyses of \cite{rabinak_early_2011} and \cite{sapir_uv/optical_2017} (hereafter \citetalias{rabinak_early_2011} and \citetalias{sapir_uv/optical_2017}) included the bound-free and bound-bound contributions to the opacity, which have a pronounced effect on the emission (\cite{nakar_early_2010} and \cite{shussman_type_2016}, hereafter \citetalias{nakar_early_2010} and \citetalias{shussman_type_2016}, included only an approximate description of the contribution to the opacity of the bound-free transitions of H).

The analytic approximations for the spherical phase of the shock cooling emission that were derived in the above mentioned papers, are valid at times during which the radiation is escaping from the outer shells of the envelope. \citetalias{sapir_uv/optical_2017} used numerical calculations of the emission from progenitors with a wide range of core/envelope mass and radius ratios to extend the validity of the analytic model of \citetalias{rabinak_early_2011} to later times, during which the emission is dominated by deeper shells where the density profile deviates from a simple power-law. 

\citetalias{shussman_type_2016} derived an analytic description for shock cooling emission including the transition between the planar and the spherical phases, based on an interpolation between the analytic solutions derived for the planar and spherical phases, and calibrated against a large set of numerical calculations. They have calculated the shock cooling emission for 120 red super-giant (RSG) progenitor structures obtained by the stellar evolutionary code MESA \citep{paxton_modules_2018}, using an effective frequency independent, i.e. "gray" opacity (neglecting the contributions of bound-bound and bound-free transitions). In addition to the analytic description, \citetalias{shussman_type_2016} argue based on analytic considerations that the description of the spectrum as a black-body with an effective color temperature $T_{\rm col}$ is not accurate at long wavelengths, and that the emission far exceeds that predicted by the Rayleigh-Jeans (RJ) tail of a Planck distribution with $T=T_{\rm col}$. These results are not supported, however, by the later analysis of \citet{kozyreva_shock_2020}, who calculated the cooling emission for 4 out of the original 120 progenitors, using the multi-group radiation-hydrodynamic code STELLA (see \citealp{blinnikov_comparative_1998,blinnikov_radiation_2000}). This later analysis finds that $T_{\rm col}$ dependence on energy during the spherical phase is in "sufficient agreement" with the analytic results of \citetalias{rabinak_early_2011} and \citetalias{nakar_early_2010} (but deviates from the results of \citetalias{shussman_type_2016}), that $L_{\rm bol}$ is overestimated by a factor of $\approx 3$ by the analytic interpolation formula, and that the long wavelength spectrum does not deviate significantly from the RJ tail of a Planck distribution (see section 4.3 \citealp{kozyreva_shock_2020}).

\subsection{Current paper}
\label{current paper}

In this paper we derive a simple analytic description of $L$ and $T_{\rm col}$ for the explosions of red supergiants (RSG) composed of compact cores surrounded by polytropic envelopes. It is valid from post breakout ($t>3R/c$), up to H recombination ($T\approx0.7$~eV) or until the photon diffusion time through the envelope becomes shorter than the dynamical time. As noted above, the shock cooling and breakout emission is only weakly dependent on the polytropic index, and \citetalias{sapir_uv/optical_2017} have shown that the shock cooling emission is not sensitive to deviations from polytropic density profiles\footnote{ \citetalias{sapir_uv/optical_2017} numerically calculated shock cooling emission for progenitors with non-polytropic density profiles that were obtained in \citet{morozova_numerical_2016} using the stellar evolutionary code MESA \citep{paxton_modules_2018}. They found $\sim10\%$ deviations in $L$ at the planar phase, and smaller deviations at the spherical phase.}. This issue is discussed further in Paper II, where we use our polytrope-based shock cooling multi-group calculation to reproduce results obtained using the radiation hydrodynamics code STELLA \citep{blinnikov_theoretical_2006} for several MESA derived progenitors with non-polytropic profiles  \citep{blinnikov_comparative_1998,tominaga_shock_2011,kozyreva_shock_2020}. We find small deviations from our analytic description, consistent with \citetalias{sapir_uv/optical_2017}.

Observations of the breakout pulse itself, $t\lesssim R/c$, may provide additional independent constraints on the progenitor and explosion parameters. Inferring $R$ from the pulse duration is not necessarily straightforward, since the duration may be dominated by the intrinsic duration of the pulse rather than by the light travel time, $R/c$, as well as by differences in the arrival time of the shock to the stellar surface at different (angular) positions due to deviations from spherical symmetry \citepalias{katz_non-relativistic_2012}. The post-breakout shock cooling phase is less sensitive to these uncertainties and may provide a direct constraint on $R$ \citepalias{katz_non-relativistic_2012,rabinak_early_2011}. Extension of the breakout pulse duration beyond $R/c$ due to strong deviations from spherical symmetry has recently been discussed by \citet{irwin_bolometric_2021}. Duration larger than $R/c$ has also been demonstrated by \citet{goldberg_shock_2022}, who carried out a 3D numerical calculation of a breakout from a convective RSG envelope. It should be pointed out that this extension is due largely to the fact that the intrinsic duration of the pulse (which is consistent with a 1D analytic estimate, see comment in \S~\ref{sec: Physics of RMS and SCE}) is significantly longer than $R/c$ for the progenitor and explosion parameters chosen in their calculation.

The analytic description derived here is based on an interpolation between the analytic results of \citetalias{sapir_non-relativistic_2011} and \citetalias{katz_non-relativistic_2012} for the planar phase of the expansion, and of \citetalias{rabinak_early_2011}/\citetalias{sapir_uv/optical_2017} for the later spherical phase, and is calibrated against the results of numerical calculations of shock cooling emission for a wide range of explosion energies,  $E=10^{50}-10^{52}$~erg, and progenitor parameters: Masses in the range $M=2-40M_{\odot}$, radii $R=3\times10^{12}-10^{14}$~cm, stellar core radii $R_{\rm c}/R=10^{-1}-10^{-3}$, envelope to core mass fractions of $M_{\rm env}/M_{\rm c}=10-0.1$, and metallicities $Z=0.1Z_{\odot}-Z_{\odot}$, where $\odot$ denotes solar values. The numerical calculations of the hydrodynamic profiles are performed assuming the plasma and radiation to be at LTE and approximating radiation transport by diffusion with constant opacity, $\kappa=0.34{\rm cm^2/g}$, corresponding to electron scattering by highly ionized 70:30 (by mass) H:He mixture. This provides a good approximation for the dynamics as long as the plasma is highly ionized, $T>0.7$~eV \citepalias{sapir_uv/optical_2017}. 

$T_{\rm col}(t)$ is obtained from the hydrodynamic profiles as the plasma temperature at the "thermal depth", from which photons diffuse out of the envelope without further absorption. The thermal depth is determined using time and space dependent effective "gray" (frequency independent) opacity, based on opacity tables that we have constructed for this purpose (and will be made publicly available), that include the contributions of bound-bound and bound-free transitions (see \S~\ref{sec:Tc}). In an accompanying paper \citepalias{morag_shock_2022-1} we show, using a large set of multi-group photon diffusion calculations, that the spectral energy distribution is well described by a Planck spectrum with $T=T_{\rm col}$, except at UV frequencies (beyond the spectral peak at $3T_{\rm col}$), where the flux is significantly suppressed due to the presence of strong line absorption. The numerical scheme used in Paper II follows that of \citet{sapir_numeric_2014}. We defer the full discussion of the multi-group results to \citetalias{morag_shock_2022-1}, where a complete descirption of the numerical scheme is given and the effects of "expansion opacity" and deviations from LTE are also discussed. For completeness, we provide here an analytic description of the UV suppression.

For the opacity we use a composite of the TOPS tables \citep{colgan_new_2018} at high temperatures and our own table produced for this work at lower temperatures. Both tables are derived under the assumption of a thermal distribution of the ionization and excitation states. The TOPS table is more suitable for high temperatures and densities, as the frequency-dependent TOPS opacities are verified against experiments at temperatures and densities exceeding tens of eV and $10^{-6} \text{g cm}^{-3}$ respectively. The line opacities in our own table are based on the \citet{kurucz_atomic_1995} line list, which is experimentally calibrated for weakly ionized species at lower temperature. Our opacity calculation code, which will be made publicly available, enables one to produce opacity tables at arbitrary $\{\rho,T,\nu\}$ resolution for convergence tests. Details of the code and tests of its validity are  discussed briefly in \S~\ref{sec:kappa} and in detail in \citetalias{morag_shock_2022-1}.

This paper is organized as follows. In \S~\ref{sec: Physics of RMS and SCE} we define our notation and summarize the analytic results of \citetalias{sapir_non-relativistic_2011}, \citetalias{katz_non-relativistic_2012}, \citetalias{rabinak_early_2011} and \citetalias{sapir_uv/optical_2017}, that we use in this paper. In \S~\ref{sec:Gray Simulations} we describe our numeric calculations, including a short discussion of our opacity tables, and present convergence tests and code verification through comparisons to known analytic solutions. Our main results are presented in \S~\ref{sec:results}: The interpolated analytic description of $L_{\rm bol}$ and $T_{\rm col}$ is given in \S~\ref{sec:MSW formula}, detailed comparisons between the analytic and numeric results are given in \S~\ref{sec:numeric_res}, and a comparison to earlier work is given in \S~\ref{sec: Previous works}. In \S~\ref{sec:gray_MG_compare} we present a brief comparison of the analytic description with the results of multi-group diffusion calculations, demonstrating the validity of the color temperature description at lower frequencies and the suppression of the UV flux due to line absorption. An analytic approximation describing the UV suppression is provided. Our results are summarized and discussed in \S~\ref{sec: Summary}.

\section{RMS breakout and Shock-Cooling Emission - summary of earlier analytic results used in this paper}
\label{sec: Physics of RMS and SCE}

At radii $r$ close to the stellar radius $R$, $\delta \equiv (R - r)/R\ll 1$, the density of a polytropic envelope approaches a power-law, 
\begin{equation}
\label{eq:rho_in}
    \rho_0 = f_\rho \bar{\rho} \delta^n.
\end{equation}
Here $\bar{\rho}\equiv M/(4\pi R^3/3)$ is the average pre-explosion density of the ejecta (exculding the mass of a possible remnant), and $n=3/2$ for convective RSG envelopes. $f_\rho$ is a numerical factor, of order unity for convective envelopes, that depends on the inner envelope structure \citep{matzner_expulsion_1999,calzavara_supernova_2004,sapir_uv/optical_2017}. The predicted breakout and cooling emission are nearly independent of $f_\rho$.

As the shock approaches the edge of the star, it accelerates down the steep density profile and the flow approaches the self-similar solutions of \citet{gandelman_shock_1956,sakurai_problem_1960}. The shock velocity diverges in this regime as
\begin{equation}
\label{eq:vs}
    v_{\rm sh} = v_{\rm s\ast} \delta^{-\beta_1 n}
\end{equation}
with $\beta_1=0.19$. $v_{\rm s\ast}$ is defined by Eq.~(\ref{eq:vs}). Based on numerical calculations, \citet{matzner_expulsion_1999} find 
\begin{equation}
\label{eq:vstar}
    v_{\rm s\ast}\approx 1.05 f_\rho^{-\beta_1}v_\ast,\quad v_\ast\equiv\sqrt{E/M},
\end{equation}
where $M$ is the mass of the ejecta, $E$ is the energy deposited in the ejecta, and $v_\ast$ is its characteristic expansion velocity. This approximation holds to better than 10\% for $M_{\rm env}/M_{\rm c}<1/3$, and overestimates $v_{\rm s\ast}$ by approx.~20\% for $M_{\rm env}/M_{\rm c}=0.1$ \citepalias[see figure 7 of ][]{sapir_uv/optical_2017}.

Breakout takes place when the optical depth of the plasma layer lying ahead of the shock equals $\tau=c/v_{\rm sh}$. We denote the shock velocity and pre-shock envelope density at this point by $\rho_{\rm bo}$ and $v_{\rm bo}$ respectively. Noting that the optical depth is dominated at this stage by Thomson scattering off the fully ionized plasma, $\tau\propto \rho_{\rm init}\delta\propto\delta^{1+n}$, we may write Eqs.~(\ref{eq:rho_in}) and~(\ref{eq:vs}) as
\begin{equation}\label{eq:rho_v_0_def}
  \rho_{\rm init}=\rho_{\rm bo} (v_{\rm bo}\tau/c)^{n/(1+n)},\quad
  v_{\rm sh} = v_{\rm bo} (v_{\rm bo}\tau/c)^{-\beta_1 n/(1+n)}.
\end{equation}
The location at which breakout "occurs", i.e. where $\tau=c/v_{\rm sh}$, is given by
\begin{equation} \label{eq:delta_bo_def}
    \delta_{\rm bo} = (n+1)\frac{c }{ \kappa \rho_{\rm bo} v_{\rm bo} R}.
\end{equation}

For RSGs, $\rho_{\rm bo}$ and $v_{\rm bo}$ are approximately related to the progenitor parameters and explosion energy by \citep{waxman_shock_2016}
\begin{align} \label{eq:rho_v_0_approx}
  \rho_{\rm bo} & = 1.16 \times 10^{-9} M_{0}^{0.32} v_{\ast,8.5}^{-0.68} R_{13}^{-1.64} \kappa_{0.34}^{-0.68} f_{\rho}^{0.45}\, \rm g \, cm^{-3}, \nonumber\\
  v_{\rm bo}/v_{\ast} & = 3.31 M_{0}^{0.13} v_{\ast, 8.5}^{0.13} R_{13}^{-0.26} \kappa_{0.34}^{0.13} f_{\rho}^{-0.09}.
\end{align}
Here, $R= 10^{13}R_{13}$~cm, $\kappa=0.34 \kappa_{0.34} \rm cm^2 g^{-1}$, and $M=1 M_{0} M_\odot$. The duration over which the breakout pulse is emitted from the star is approximately given by the shock crossing time of the breakout layer,
\begin{equation}
\label{eq:tbo}
   \frac{\delta_{\rm bo}R}{v_{\rm bo}} =\frac{(n+1)c }{ \kappa \rho_{\rm bo} v_{\rm bo}^2}= (n+1)t_{\rm bo}=74.9\, \rho_{\rm bo,-9}^{-1}\kappa_{0.34}^{-1}v_{\rm bo,9}^{-2}{\, \rm s},
\end{equation}
where $\rho_{\rm bo}=10^{-9}\rho_{\rm bo,-9}{\rm \, g \, cm^{-3}}$ and $v_{\rm bo}= 10^{9} v_{\rm bo,9}{\rm cm/s}$. The observed pulse duration may be longer than this intrinsic duration due to light travel time effects, which spread the pulse over $R/c$\footnote{For the parameters used in the 3D calculation of \citet{goldberg_shock_2022}, $R=820R_\odot$, $M=12.7M_\odot$ and $E=0.8\times10^{51}$~erg, the breakout density, velocity and duration given by Eqs.~(\ref{eq:rho_v_0_approx}) \&~(\ref{eq:tbo}) are $2\times10^{-10}{\rm g/cm^3}$, $5\times10^3{\rm km/s}$ and $(n+1)t_{\rm bo}=4300\,{\rm s}>R/c=1900$~s respectively. These 1D analytic estimates are consistent with the results obtained in the 3D calculation, $1.5-3\times10^{-10}{\rm g/cm^3}$, $3-5\times10^3{\rm km/s}$ and $\approx1.6$~hr rise time (The 1D analytic results and the 3D numeric results differ from the 1D calculations based on MESA profiles used in \citet{goldberg_shock_2022}, since the 1D profiles reach down only to a density of $1\times10^{-9}{\rm g/cm^3}$, that is, the 1D profiles do not resolve the breakout shell). The pulse duration is longer than $R/c$ since the shock crossing time of the breakout layer is larger than $R/c$, independent of 3D effects. We note that the breakout shell is not well resolved in the 3D calculations of \citet{goldberg_shock_2022}- For the parameters chosen, the relative thickness of the breakout shell, Eq.~(\ref{eq:delta_bo_def}), is $\delta_{\rm bo}=0.035$ while the radial resolution in the numeric calculation is $\delta R/R=0.01$. Thus, higher resolution calculations are required in order to draw quantitative conclusions regarding the breakout properties from the 3D calculations. 
}.
$\delta_{\rm bo}$ is given as a function of progenitor parameters and explosion energy as
\begin{equation}
\label{eq:dbo}
    \delta_{\rm bo}=0.02 \, R_{13}^{0.90} (f_\rho M_0 \, v_{s*,8.5}\, \kappa_{0.34})^{-0.45}.
\end{equation}

\citetalias{sapir_non-relativistic_2011}, \citetalias{katz_non-relativistic_2012} and \citetalias{sapir_non-relativistic_2013} provide an exact numeric solution, and approximate analytic solutions, for the breakout emission and for the shock-cooling emission during the planar phase of the expansion. In this paper we are interested in developing a simple analytic expression for the shock cooling emission following the shock breakout pulse at
\begin{equation}\label{t_lc}
  t > 3R/c=17\,R_{13}{\rm \, min},
\end{equation}
for which the finite travel time effect is small. At this phase, the exact solution is well approximated by (\citetalias{katz_non-relativistic_2012})
\begin{align}
\label{eq: L_planar_powerlaw}
    L_{\rm planar} & =3.21\times10^{42} \, R_{13}^{2}\left(v_{\rm bo,9}\rho_{\rm bo,-9}^{-1}t_{\rm hr}^{-4}\kappa_{0.34}^{-4}\right)^{1/3}\, {\rm erg\,s}^{-1} 
    \nonumber\\
    & = 3.01 \times 10^{42} \, R_{13}^{2.46} v_{\rm s* , 8.5}^{0.60} (f_{\rho}M_0)^{-0.06} t_{\rm hr}^{-4/3} \kappa_{0.34}^{-1.06}  \, \rm erg \, s^{-1}
\end{align}
and
\begin{align}
\label{eq: T_planar_powerlaw}
    T_{\rm ph, planar} & = \left( L/4\pi\sigma R^2 \right)^{1/4} \nonumber\\
      & = 7.06\,v_{\rm bo,9}^{1/12}\rho_{\rm bo,-9}^{-1/12}t_{\rm hr}^{-1/3}\kappa_{0.34}^{-1/3}\, \rm eV \nonumber\\
      & = 6.95\,R_{13}^{0.12} v_{\rm s*,8.5}^{0.15}(f_{\rho}M_0)^{-0.02} t_{\rm hr}^{-1/3}\kappa_{0.34}^{-0.27}\, \rm eV\,.
\end{align}
Here, $t=1 \, t_{\rm hr} \, \rm hour$ and we define $t=0$ as the time at which the breakout flux peaks.

The \citetalias{rabinak_early_2011} approximations for $L$ and $T_{\rm col}$ during the spherical phase can be recast in terms of the breakout density $\rho_{\rm bo}$ and velocity $v_{\rm bo}$ using Eqs.~(\ref{eq:rho_in}), (\ref{eq:vs}) and (\ref{eq:rho_v_0_def}) as
\begin{align}
\label{eq:LRW}
    L_{\rm RW} & =2.24\times10^{42} \, R_{13}^{0.84}v_{\rm bo,9}^{1.49}\rho_{\rm bo,-9}^{-1/3}\kappa_{0.34}^{-4/3} t_{\rm d}^{-0.17}\,
    {\rm erg\,s}^{-1} \nonumber\\
    &=2.08\times10^{42} \, R_{13} v_{\rm s*,8.5}^{1.91}(f_{\rho}M_0)^{0.09}\kappa_{0.34}^{-0.91} t_{\rm d}^{-0.17}\,
    {\rm erg\,s}^{-1},
\end{align}
and (see also \citetalias{sapir_uv/optical_2017})
\begin{equation}
\label{eq:TcRW}
    T_{\rm col, RW}=f_{T}T_{\rm ph,RW},\quad f_T=1.1,
\end{equation}
where the photospheric temperature (defined as the temperature at the location $\tau=1$ for Thomson opacity) is
\begin{align}
\label{eq:TphRW}
    T_{\rm ph,RW} &=1.69\,R_{13}^{0.12}v_{\rm bo,9}^{0.01}\rho_{\rm bo,-9}^{-1/12}\kappa_{0.34}^{-1/3} t_{\rm d}^{-0.45}{\rm \, eV}  \nonumber\\
    &=1.66\,R_{13}^{1/4} v_{\rm s*,8.5}^{0.07}(f_{\rho}M_0)^{-0.03}\kappa_{0.34}^{-0.28} t_{\rm d}^{-0.45}\,{\rm \, eV}.
\end{align}
These spherical phase approximations hold at
\begin{equation}
\label{eq:tranges}
    \max[t_{\rm exp},t_{\rm ph}]<t<\min[t_{0.7 \rm eV},t_\delta],
\end{equation}
where 
\begin{equation} \label{eq:t_hom}
    t_{\rm exp}= R/5v_{\rm s \ast} = 1.76 \, R_{13} v_{\rm s*,8.5}^{-1} \, \rm hours
\end{equation}
is the time required for significant expansion of the emitting shells, 
\begin{equation}\label{eq:t_pbo}
  t_{\rm ph}= 6.43 \, R_{13}^{1.39} v_{\rm s*,8.5}^{-1.69} (f_\rho M_0\kappa_{0.34})^{-0.19} \rm \, hours
\end{equation}
is the time required for the photosphere to penetrate beyond the thickness of the breakout shell,
\begin{equation}\label{eq:t_deep}
  t_\delta= 3.19 \, v_{\rm s*,8.5}^{-1} f_\rho^{-0.11} \sqrt{ M_0\kappa_{0.34}} \rm \, days
\end{equation}
is the time at which the photosphere penetrates to a depth of $10^{-2.5}$ of the progenitor mass (corresponding to $\delta\sim0.1$), where the density profile is no longer described by Eq.~(\ref{eq:rho_in}), and
\begin{equation} \label{eq:t_0_7eV}
    t_{0.7 \rm eV} = 6.86 \, R_{13}^{0.56} v_{\rm s*,8.5}^{0.16} \kappa_{0.34}^{-0.61} (f_{\rho}M_0)^{-0.06} \rm  \, days
\end{equation}
is the time at which the temperature drops to $0.7$~eV. The resulting spectral energy distribution is then defined using the blackbody intensity $B_\nu$ as follows,

\begin{equation}
    L_{\nu}=L\times\pi B_{\nu}(T_{\rm col})/\sigma T_{\rm col}^{4},
    \label{eq:L_nu_BB_formula}
\end{equation}
where $\sigma$ is the blackbody constant.

\citetalias{sapir_uv/optical_2017} extended the validity time of the \citetalias{rabinak_early_2011} model beyond $t_\delta$, including an analytic approximation for the suppression of the luminosity below the value given by Eq.~(\ref{eq:LRW}), based on the results of numerical calculations. They find that the flux is suppressed by a factor
\begin{equation}\label{eq:L_SW17}
  L/L_{\rm RW}= A\exp\left[-\left(\frac{at}{t_{\rm tr}}\right)^{\alpha}\right]
\end{equation}
with $\{A,a,\alpha\}=\{0.94,1.67,0.8\}$. The transparency time $t_{\rm tr}$, at which photons escape from deep in the envelope over a dynamical time, is given by
\begin{equation}
\label{eq:t_transp}
    \begin{split}
        \begin{aligned}
            t_{\rm tr} &= \sqrt{\frac{\kappa M_{\rm env}} {8 \pi c v_{\rm s \ast} }}  \\ &= 19.5 \, \sqrt{M_{\rm env,0} \kappa_{0.34} v_{\rm s*,8.5}^{-1}} \, \text{days},
        \end{aligned}
    \end{split}
\end{equation}
where $M_{\rm env}= M_{\rm env,0} M_\odot$ is the mass of the envelope. This modification extends the model validity time $t_{\delta}$ to $t\approx t_{\rm tr}/a$ (see Eq.~(\ref{eq:tranges})).

\section{Description of the numerical code}
\label{sec:Gray Simulations}

We solve numerically the radiation hydrodynamics equations of a spherically symmetric flow of an ideal fluid with radiation dominated pressure, assuming LTE (of the plasma and the radiation) and approximating radiation transport by diffusion with constant opacity $\kappa=0.34{\rm cm^2/g}$, corresponding to electron scattering by highly ionized 70:30 (by mass) H:He mixture. The equations and boundary conditions are given in \S~\ref{sec:eqs}, and the initial conditions are described in \S~\ref{sec:init}. The determination of the thermal depth and of $T_{\rm col}(t)$, based on the numerically derived hydrodynamic profiles, is described in \S~\ref{sec:Tc}. Our opacity tables, that include the contributions of bound-bound and bound-free transitions and are used for the determination of the thermal depth, are briefly described in \S~\ref{sec:kappa}. The validation of the numeric code and the convergence of the calculations are described in \S~\ref{sec:valid}.

\subsection{Radiation-hydrodynamics equations}
\label{sec:eqs}

We solve numerically the (Lagrangian) radiation-hydrodynamics equations in spherical symmetry under the diffusion approximation, 
\begin{equation}
\frac{dr}{dt}=v\label{eq:matter continuity-1},
\end{equation}
\begin{equation}
    \rho = \rho_{0} \frac{r_{0}^2}{r^2} \frac{\partial r_0} {\partial r},
\end{equation}
\begin{equation}
\frac{dv}{dt}=-\frac{1}{\rho}\frac{1}{r^{2}}\frac{d}{dr}\left(\frac{1}{3}r^{2}u\right)\label{eq: momentum continuity},
\end{equation}
\begin{equation}
\frac{du}{dt}=\left.\frac{\partial u}{\partial t}\right|_{\rm compression}+\left.\frac{\partial u}{\partial t}\right|_{\rm diffusion},
\end{equation}
\begin{equation}
\left.\frac{\partial u}{\partial t}\right|_{\rm compression}=-\frac{4}{3}\frac{u}{r^{2}}\frac{\partial\left(r^{2}v\right)}{\partial r},
\end{equation}
\begin{equation}
\left.\frac{\partial u}{\partial t}\right|_{\rm diffusion}=-\frac{1}{r^2}\frac{\partial (r^2j)}{\partial r},
\end{equation}
where the photon energy flux density $j$ is given by
\begin{equation}
\label{eq:flux}
j=\frac{-1}{\rho\kappa_{\rm es}}\left(\frac{c}{3}\frac{\partial u}{\partial r}-\frac{1}{c}\frac{\partial j}{\partial t}\right).
\end{equation}
$r,\rho,v$, $u$ are the radius, density, velocity and radiation energy density of the fluid elements, and $0$ subscripts denote initial values. $\kappa_{\rm es}$ is the Thomson scattering opacity. Each of these equations is solved explicitly using operator splitting, with the exception of the diffusion term, where $u$ and $j$ are solved for together implicitly. The flux is initially started as $j_0=\frac{-1}{\rho\kappa_{\rm es}}\frac{c}{3}\frac{\partial u_0}{\partial r}$. The time derivative term in eq.~(\ref{eq:flux}) becomes important in optically thin regions. In our calculations it improves numerical issues, but does not have an important effect on flux since the luminosity is determined deep in the ejecta. We verified this by producing simulations with and without the time derivative term for progenitor radii $R=3\times10^{12}$ cm, $10^{13}$, cm, and $10^{14}$ cm (explosion energies $E=10^{51}$ erg and core, envelope masses of $M_{\rm c}=M_{\rm e}=10M_\odot$), finding $<1\%$ effect on $L$ and $T_{\rm col}$ during shock-cooling.

\subsection{Initial and boundary conditions}
\label{sec:init}

Following \citetalias{sapir_uv/optical_2017}, we adopt a simplified description of the initial stellar structure: A uniform density core of radius $R_{\rm c}$ and mass $M_{\rm c}$, surrounded by a polytropic envelope of mass $M_{\rm env}$ and outer radius $R$ at hydrostatic equilibrium. The explosion is initiated by depositing energy $E$ at the core, uniformly distributed within the few innermost cells \footnote{The total energy in the hydrodynamic simulations at late times matches the injected energy to $0.3\%$}. We assume a static reflective boundary in the inner surface, and a free boundary at the outer surface that accelerates as $\partial_t v_{\rm b} = j_{\rm b}\kappa / c$, where the subscript b denotes boundary values. The boundary flux is given by  $j_{\rm b}=f_{\rm edd}cu_{\rm b}$, where $f_{\rm edd}=0.3-0.5$ is the Eddington factor. The results are insensitive to the exact value of$f_{\rm edd}$ since the flux is determined deep within the plasma, at $\tau\sim c/v\gg1$.

Numerical calculations were performed for a wide range of explosion energies,  $E=10^{50}-10^{52}$~erg, and progenitor parameters: masses in the range $M=2-40M_{\odot}$, radii $R=3\times10^{12}-10^{14}$~cm, stellar core radii of $R_{\rm c}/R=10^{-1}-10^{-3}$ envelope to core mass fractions of $M_{\rm env}/M_{\rm c}=10-0.1$, and metallicities $Z=0.1Z_{\odot}-Z_{\odot}$. The quantities $f_{\rho}M$ and $v_{\rm s,*}$ vary in our simulations between $2-45 \, M_\odot$ and $6\times10^7-6\times10^8 \, \rm cm \, s^{-1}$.

In setting up the initial hydrostatic equilibrium density profile, we neglect the mass of a possible remnant (in our calculations, the entire mass, $M=M_{\rm c}+M_{\rm env}$, is ejected). This implies that the density profile differs somewhat from that of an explosion with finite remnant mass $M_{\rm rem}$. This difference has, however, only a negligible effect on the results. This can be seen by noting that the inclusion of $M_{\rm rem}$ implies a modification of the outer density profile through a modification of $f_\rho$ in Eq.~(\ref{eq:rho_in}). However, the breakout and cooling emission are almost independent of $f_\rho$, see Eqs.~(\ref{eq: L_planar_powerlaw}-\ref{eq:TphRW}). In addition, the dependence of $f_\rho$ on $M_{\rm rem}$ is also not strong, $f_{\rho}\approx[M_{\rm env}/(M_{\rm c}+M_{\rm rem})]^{1/2}$ \citepalias{sapir_uv/optical_2017}.

The numerical calculation is carried out in two steps. First, a hydrodynamics only calculation is carried out, not including radiation diffusion and capturing the shock wave using artificial viscosity. Then, a radiation-hydrodynamics calculation is initiated with the hydrodynamic profiles obtained numerically at a time preceding the arrival time of the shock at the surface by $24t_{\rm bo}$. We modify the hydrodynamic profiles used for the initiation of the radiation hydrodynamics profile by "smearing" the shock structure obtained using artificial viscosity to match the "Sakurai-Weaver Anzats" structure of \citetalias{sapir_non-relativistic_2011}, which provides an excellent approximation to the hydrodynamic profiles obtained in radiation-hydrodynamics calculations under the diffusion approximation. This modification enables the shock to converge on the final RMS shape in fewer shock crossing times.

\subsection[Determining Tcol]{Determining $T_{\rm col}$}
\label{sec:Tc}
$T_{\rm col}(t)$ is obtained from the hydrodynamic profiles as the plasma temperature at the "thermal depth", $r_{\rm Th}(t)$, from which photons diffuse out of the envelope without further absorption. Following \citetalias{rabinak_early_2011} and \citetalias{sapir_uv/optical_2017}, we define $r_{\rm Th}(t)$ by 
\begin{equation}
\label{eq:Itais_Prescription}
\int_{r_{\rm Th}(t)}\rho\sqrt{3\kappa_{\rm R}\left(\kappa_{\rm R}-\kappa_{\rm es}\right)}dr'=1.
\end{equation}
Here, $\kappa_{\rm es}(\rho,T)$ is the electron scattering opacity, accounting for the ionization fraction, $\kappa_{\rm R}(\rho,T)$ is Rosseland mean opacity, calculated from the high frequency resolution opacity tables (see \S~\ref{sec:kappa}), and $(\kappa_{\rm R}-\kappa_{\rm es})$ represents the effective absorption opacity, $\kappa_{\rm abs}$.

\subsection{Our composite opacity table}
\label{sec:kappa}

In order to extract $T_{\rm col}$ (see Eq.~(\ref{eq:Itais_Prescription})) we use a composite of the TOPS opacity tables \citep{colgan_new_2018} and our own tables, which include the contributions of free-free, bound-free and bound-bound transitions. The bound-bound contribution in our table is based on the Kurucz database of atomic transition lines \citep{kurucz_atomic_1995}. Thermal distributions of ionization and excitation states are assumed. The TOPS table is more suitable for high temperatures and densities, as the frequency-dependent TOPS opacities are verified against experiments at temperatures and densities exceeding tens of eV and $10^{-6} \text{g cm}^{-3}$ respectively \citep{colgan_light_2015,colgan_new_2016,colgan_new_2018}. The Kurucz data base of atomic transitions was calibrated against measured line frequencies and oscillator strengths \citep{kurucz_atomic_1995}, but is incomplete for highly ionized species at high, $T>10$~eV, temperature.

\begin{figure}
    \centering
    \includegraphics[width = \columnwidth]{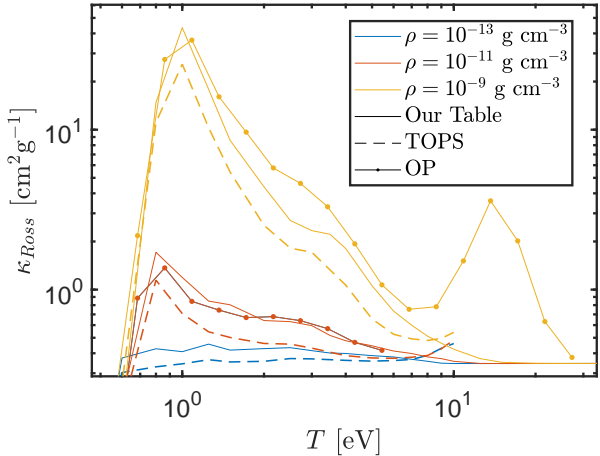}
    \caption{Rosseland mean opacity for a solar mix of atoms as a function of density $\rho$ and temperature $T$. The opacity obtained using our table is compared with those obtained using the TOPS and OP tables. For deriving $T_{\rm col}$ from the numeric calculations we use the TOPS Rosseland mean above 4~eV, and our Rosseland mean below. See discussion in \S~\ref{sec:kappa}.}
    \label{fig:SMRossCompare}
\end{figure}

Our table and the TOPS table both provide a frequency-dependent opacity, that we use in our multi-group calculations (\citetalias{morag_shock_2022-1}). For the calculation of $T_{\rm col}$ in this paper we use only the Rosseland mean opacity. A comparison between the Rosseland mean opacity obtained from our table and that obtained from the publicly available TOPS \citep{colgan_new_2018} and OP \citep{iglesias_updated_1996} tables is shown in Fig. \ref{fig:SMRossCompare} for a mixture of ten atoms representing a solar mix, focusing on the temperature regime of 1 - 10~eV. At temperatures below 10~eV, the resulting opacities are comparable, with our results lying roughly between the TOPS and OP values. Above 10~eV, TOPS and OP show a clear opacity bump due to atomic lines that are not included in our table (see preceding paragraph).

Comparing $T_{\rm col}$ obtained from Eq.~(\ref{eq:Itais_Prescription}) using our and the TOPS opacity tables separately, we find in general differences of tens of percents in the derived temperatures (see \S~\ref{sec:numeric_res}). Much better agreement is obtained when the photosphere temperature, given by Eq.~(\ref{eq:TphRW}), is approximately 4~eV. For deriving $T_{\rm col}$ we therefore use the TOPS Rosseland mean above 4~eV, and our Rosseland mean below. 

\subsection{Code validation and numerical convergence}
\label{sec:valid}

We validated our code by comparing our hydrodynamics-only calculations of shock breakout to the analytic self-similar solutions of \citet{gandelman_shock_1956} and \citet{sakurai_problem_1960}, and by comparing the radiation hydrodynamics calculations of shock breakout to the analytic "Sakurai-Weaver Anzats" solutions of \citetalias{sapir_non-relativistic_2011} for both hydrodynamics ($\rho$,$v$, and $p$) and luminosity. We found excellent agreement between the numeric and analytic results.

The hydrodynamic only results shown below were obtained by calculations with $8000$ grid cells. Comparing the results obtained with this resolution to results obtained with $4000$ grid cells, we find that the velocity, density and pressure at $t=R/v_{\rm bo}$ are converged to better than $0.2\%,2\%,\text{ and }3\%$ respectively. The radiation hydrodynamics results shown below were obtained by calculations with $3200$ grid cells. The initial cell grid for the gray simulations is a smoothed function, with modest resolution in the interior, highest resolution at the starting location of the shock, and steadily decreasing resolution outwards \citep[keeping cell count constant across the RMS - see ][]{sapir_uv/optical_2017}, before approaching a constant resolution for $\tau\lesssim c/v_{\rm bo}$. The scattering optical depth of the outermost cell is a few percent. Comparing the results obtained with this resolution to results obtained with $1600$ cells, we find that the velocity, density, and pressure are converged to better than $0.5\%,2\%,2\%$ respectively, and that $L$ is converged to better than $3\%$.

The post-breakout velocity is increasing throughout most of the ejecta, and nearly uniform at the breakout shell. The velocity profile obtained in the numeric calculations shows a small decrease, of approximately 0.1\%, across the breakout shell. While the decrease is small, the positive velocity divergence leads after large expansion to the formation of a high density shell at the outer part of the breakout layer (high density compared to the adjacent parts of the ejecta). This effect has been observed in earlier work \citep{falk_radiation_1977,ensman_shock_1992,hauschildt_analysis_1994,blinnikov_comparative_1998}. The positive density gradient is expected to lead to the development of Rayleigh-Taylor instability, which is not described by our code, and the optical depth of the high density shell is low, $<10^{-2}$, so that it does not affect our results. We have therefore introduced into our numeric calculation an adiabatic plasma energy term $e$ that is coupled to an artificial viscosity $q$, modifying Eq.~\ref{eq: momentum continuity} to
\begin{equation}
\frac{dv}{dt}=-\frac{1}{\rho}\frac{1}{r^{2}}\frac{d}{dr}\left(r^{2}[(\gamma-1)e+\frac{1}{3}u+q]\right),\label{eq: momentum continuity with e q}
\end{equation}
with $\gamma=5/3$. $e$ is initially set to zero and evolves according to
\begin{equation}
\frac{\partial e}{\partial t}=-\frac{1}{r^{2}}\left(\gamma e+ q\right)\frac{\partial\left(r^{2}v\right)}{\partial r},
\end{equation} \label{eq: e compression}
$e$ and $q$ are largest at the high-density shell, but smaller than the radiation density by orders of magnitude. Their addition circumvents numeric problems due to the strong positive density gradient produced by the slight decrease in the velocity profile.

\subsubsection{Opacity convergence for the calculation of $T_{\rm col}$}
\label{sec: kappa converge}

The Rosseland mean opacity is calculated over a grid of values of $T$ and $R\equiv T^{3}/\rho$. $R$ is binned logarithmically and $T$ nearly logarithmically, with higher $T$ resolution below 1~eV, in the region near recombination\footnote{Namely $T=\text{10}^{x^{2}+\alpha_{min}}$~eV, where $x$ is distributed linearly between $0<x<\sqrt{\alpha_{max}-\alpha_{min}}$, which gives $10^{\alpha_{min}}<T<10^{\alpha_{max}}$.}.
We checked the convergence of the derived $T_{\rm col}$ with respect to our table's underlying $\{R,T,\nu\}$ grid, where $\nu$ is integrated over to produce the Rosseland mean. $T_{\rm col}$ deviates by less than $5\times10^{-3}$ when comparing results using grid size $\{N_{R},N_{T}\}=\{30,120\}$ against $\{N_{R},N_{T}\}=\{16,66\}$  for \{$R$,$T$\}. The fractional deviation in $\kappa_{\rm R}$ is less than $5\times 10^{-3}$ between a table with resolution of $\Delta\nu/\nu\sim10^{-5}$ and $\Delta\nu/\nu\sim10^{-6}$ \footnote{It is not practical to produce an entire Rosseland mean table for $\Delta\nu/\nu\sim10^{-6}$ resolution. Therefore, we check convergence with this resolution only for regions of parameters for which the convergence was not sufficient when comparing results with $\Delta \nu / \nu \sim 10^{-4}$ and $10^{-5}$.}.

The limited temperature resolution offered by TOPS makes testing convergence with respect to temperature resolution difficult. However, the TOPS $\rho$ and $T$ resolution is comparable to our $\{N_{R},N_{T}\}=16\times66$ grid, suggesting that the results are similarly converged.
Finally, we note that TOPS limit their table output to densities exceeding $10^{-12}{\rm g\,cm^{-3}}$. This is higher than the density obtained in our simulations at late times. However, $T_{\rm col}$ is not sensitive to the opacity at such densities for $T_{\rm ph}>4$~eV.

\section{Results}
\label{sec:results}

We first describe in \S~\ref{sec:MSW formula} our calibrated analytic interpolation equations, and then present in \S~\ref{sec:numeric_res} comparisons of our numeric results to the analytic description. Our results are compared to those of earlier works in \S~\ref{sec: Previous works}.

\subsection{Calibrated analytic model}
\label{sec:MSW formula}

The numerically calculated $L$ and $T_{\rm col}$ are well approximated by the following interpolations between the \citetalias{katz_non-relativistic_2012} solution for the planar phase, given by Eqs.~(\ref{eq: L_planar_powerlaw}) and~(\ref{eq: T_planar_powerlaw}), and the \citetalias{rabinak_early_2011}/\citetalias{sapir_uv/optical_2017} approximate solution for the spherical phase, given by Eqs.~(\ref{eq:LRW}-\ref{eq:TphRW}), (\ref{eq:L_SW17}) and~(\ref{eq:t_transp}):
\begin{equation}
\label{eq:L_sum}
    L=L_{\rm planar}+A\exp\left[-\left(\frac{at}{t_{\rm tr}}\right)^{\alpha}\right]\,L_{\rm RW},
\end{equation}
\begin{equation}
\label{eq:T_col_MSW_2}
    T_{\rm col}=f_{T}\min\left[T_{\rm ph,planar}\,,\,T_{\rm ph,RW}\right],\quad f_T=1.1.
\end{equation}
We find that $\{A,a,\alpha\}=\{0.9,2,0.5\}$ provide a better overall agreement to the numeric results compared to $\{A,a,\alpha\}=\{0.94,1.67,0.8\}$ used in \citetalias{sapir_uv/optical_2017}.

These approximations hold at
\begin{equation}
\label{eq:t_lc1}
   3 R / c = 17\,R_{13} \, {\rm min} < t < \min[t_{\rm 0.7 eV}, t_{\rm tr}/a].
\end{equation}
The lower limit is set by the requirement that light travel time effects be negligible (as confirmed numerically, see \S~\ref{sec:numeric_res}), and the upper limits are set by the time at which the temperature drops to 0.7~eV, Eq. (\ref{eq:t_0_7eV}), and the time at which the photon escape time from deep within the envelope becomes comparable to the dynamical time, Eq. (\ref{eq:t_transp}).

The analytic model can be rewritten more simply by normalizing the time to $t=t_{\rm br}$, the time at which $L_{\rm planar}=L_{\rm RW}$. Defining
\begin{eqnarray}\label{eq:trans_def}
  L_{\rm br}&=&L_{\rm planar}(t=t_{\rm br})=L_{\rm RW}(t=t_{\rm br}),\nonumber\\
  T_{\rm col,br}&=&f_T T_{\rm ph,RW}(t=t_{\rm br}),\nonumber\\
  \tilde{t}&=&t/t_{\rm br},
\end{eqnarray}
we have
\begin{equation}
\label{eq:L_trans}
    L/L_{\rm br}=\tilde{t}^{-4/3}+ A\exp\left[-\left(at/t_{\rm tr}\right)^{\alpha}\right]
    \tilde{t}^{-0.17},
\end{equation}
and
\begin{equation}
\label{eq:T_trans_vbo_factor}
    T_{\rm col}/T_{\rm col,br}=\min\left[0.98 \, v_{\rm bo,9}^{-0.05}\tilde{t}^{-1/3},\tilde{t}^{-0.45}\right].
\end{equation}
The coefficient $0.98 \, v_{\rm bo,9}^{-0.05}$ lies in the range $[0.95 -1]$ and is a result of the fact that $t_{\rm br}$ does not exactly correspond to the time when $T_{\rm planar} = T_{\rm RW}$. For simplicity, we use
\begin{equation}
\label{eq:T_trans}
    T_{\rm col}/T_{\rm col,br}=\min\left[0.97\,\tilde{t}^{-1/3},\tilde{t}^{-0.45}\right],
\end{equation}
when comparing to our numeric results. This change does not affect the final quoted systematic error.

The break values are quantities that, along with $t_{\rm tr}$, can be directly deduced from observations. They are given in terms of the model parameters $v_{\rm s*}$\,,$f_{\rho}M$, and $R$ as
\begin{equation}
\label{eq:t_br_of_vs}
    t_{\rm br}= 0.86 \, R_{13}^{1.26} v_{\rm s*,8.5}^{-1.13}
(f_{\rho}M_0\kappa_{0.34})^{-0.13}\,\text{hrs},
\end{equation}
\begin{equation}
\label{eq:L_br_of_vs}
L_{\rm br}=3.69\times10^{42} \, R_{13}^{0.78} v_{\rm s*,8.5}^{2.11}
(f_{\rho}M_0)^{0.11} \kappa_{0.34}^{-0.89}\,{\rm erg \, s^{-1}},
\end{equation}
\begin{equation}
\label{eq:T_br_of_vs}
T_{\rm col,br}= 8.19 \, R_{13}^{-0.32} v_{\rm s*,8.5}^{0.58}
(f_{\rho}M_0)^{0.03} \kappa_{0.34}^{-0.22}\,{\rm eV}.
\end{equation}

\subsection{Numeric results}
\label{sec:numeric_res}

We present in this section a comparison of the numeric results with the analytic description given in \S~\ref{sec:MSW formula}. For each numeric calculation, we determined $\rho_{\rm bo}$ and $v_{\rm bo}$ by fitting the numeric shock breakout luminosity pulse of duration $\sim R/c$ to the numeric results given in table 3 of \citetalias{sapir_non-relativistic_2011}, and then used these values for obtaining the analytic model flux using Eqs.~(\ref{eq: L_planar_powerlaw}), (\ref{eq: T_planar_powerlaw}),~(\ref{eq:LRW}-\ref{eq:TphRW}). This provides a better accuracy for the analytic flux compared to that obtained by inferring $v_{\rm s\ast}$ directly from the shock velocity in the corresponding hydrodynamic-only run. 

\begin{figure}
    \includegraphics[width = \columnwidth]{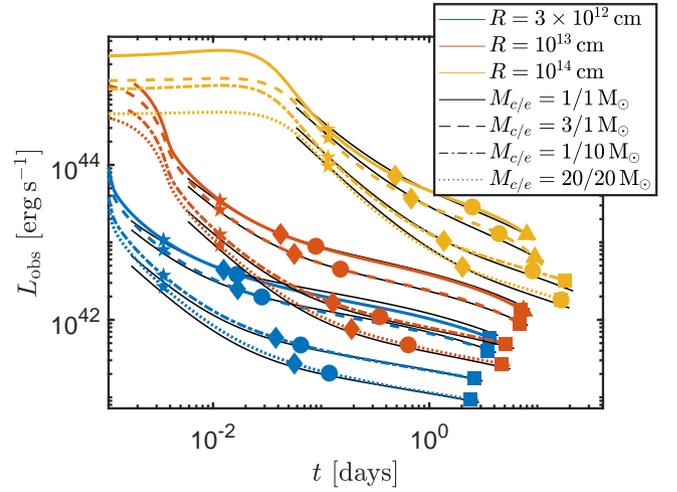}
    \caption[LMSW LTT caption]{Numerically derived bolometric luminosities shown in color for different progenitor radii and mass combinations ($M_{c/e}$ denote core and envelope masses) and explosion energy of $E=10^{51}$ erg, compared to our analytic results, Eq.~(\ref{eq:L_sum}) in black. The numeric results shown include light-travel time effects\footnotemark,
    which modify $L$ at $t\lesssim R/c$. The symbols indicate the various validity times: $\star$ corresponds to $3R/c$ ; $\blacklozenge$ to $t_{\rm exp}$ (Eq. \ref{eq:t_hom}); $\bullet$ to $t_{\rm ph}$ (Eq. \ref{eq:t_pbo});  $\blacktriangle$ to $t_{\rm tr}/a$ (Eq. \ref{eq:t_transp}); $\blacksquare$ to $t_{\rm 0.7 eV}$ (Eq. \ref{eq:t_0_7eV})}
    \label{fig:L_MSW}
\end{figure}
\footnotetext{We calculate these effects using eq. 30 of \citetalias{katz_non-relativistic_2012}. The time $t'$, at which a photon that is observed at time $t$ left the photosphere, located at $r_{ph} (t')$, is given self consistently by $t'=t+\left(\mu\,r_{ph}(t')-R\right)/c$. $r_{ph} (t')$ is inferred from the numeric solution, improving on the KSW II approximation of $r_{ph}=R$.}

\begin{figure}
    \includegraphics[width = \columnwidth]{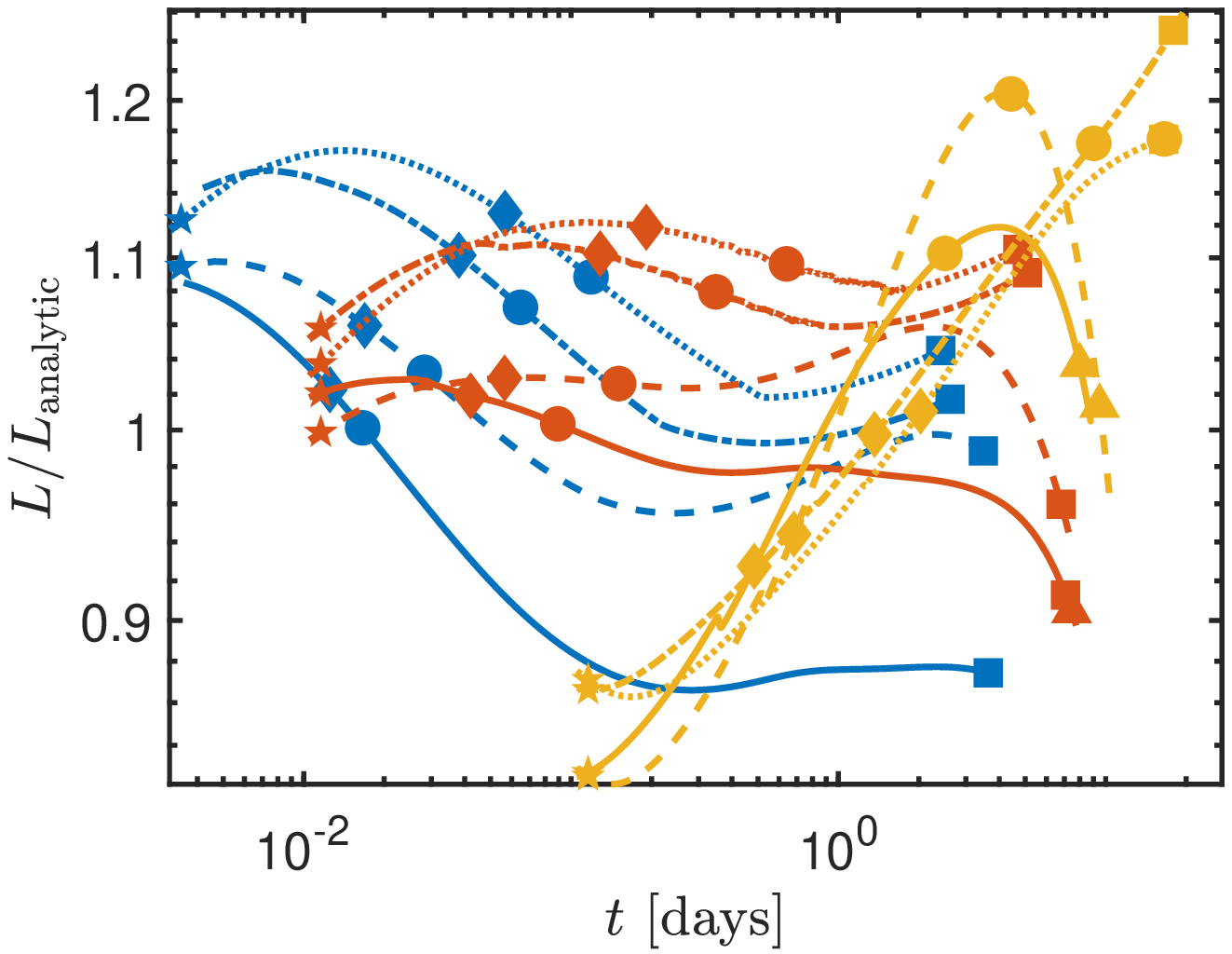}
    \includegraphics[width = \columnwidth]{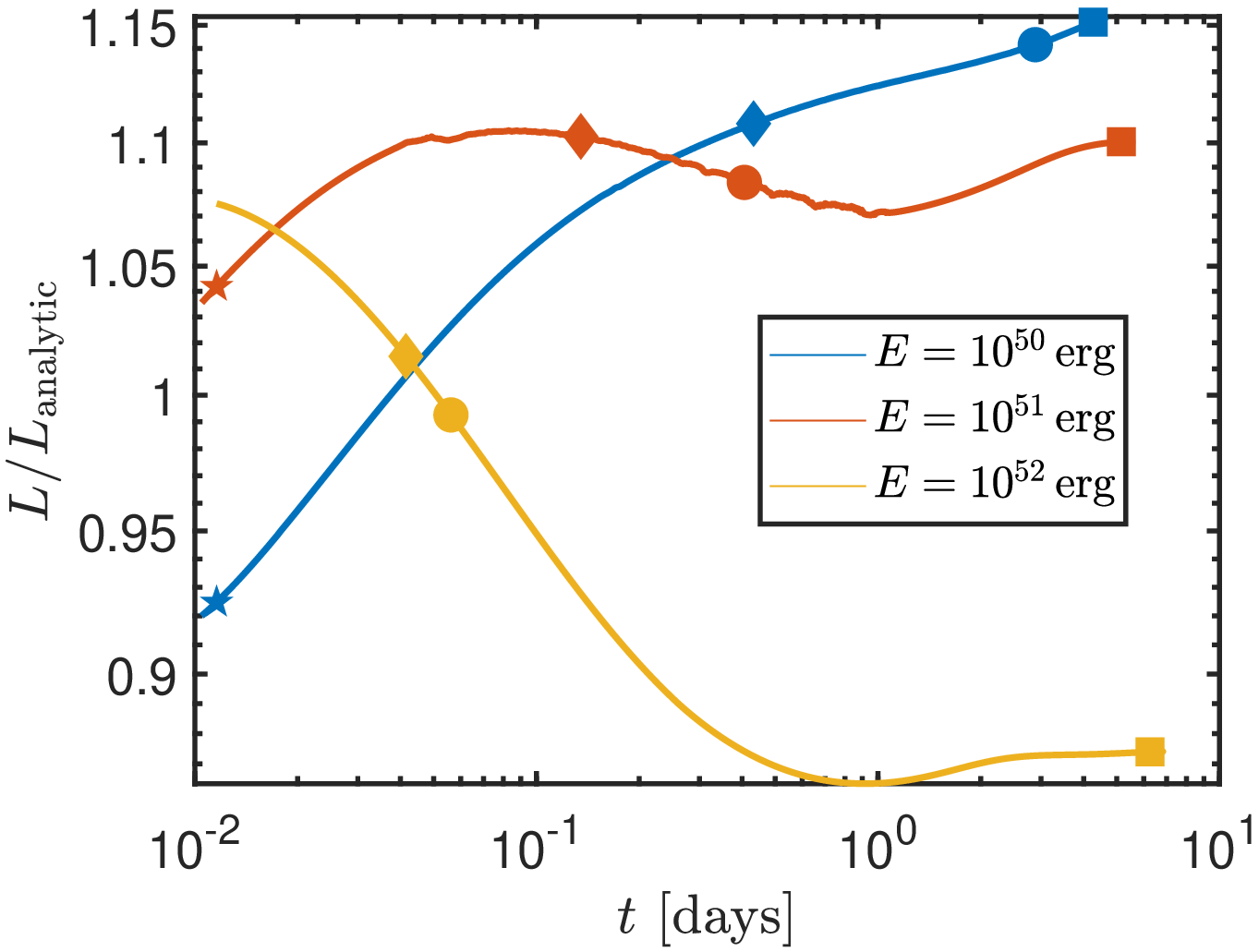}

    \caption{\label{fig:Eser L} The ratio of numerically derived bolometric luminosities and the analytic results of Eq.~(\ref{eq:L_sum}). Top: Line colors and styles correspond to the same values of parameters as in Fig.~\ref{fig:L_MSW}. Bottom: Results obtained for a range of explosion energies and $R=10^{13} \, \rm cm$, $M_{\rm c}=M_{\rm env}=10M_\odot$. Marks indicate the same validity times as in Fig. \ref{fig:L_MSW}.}
    \label{fig:L_over_analytic}
\end{figure}

\begin{figure}
    \includegraphics[width = \columnwidth]{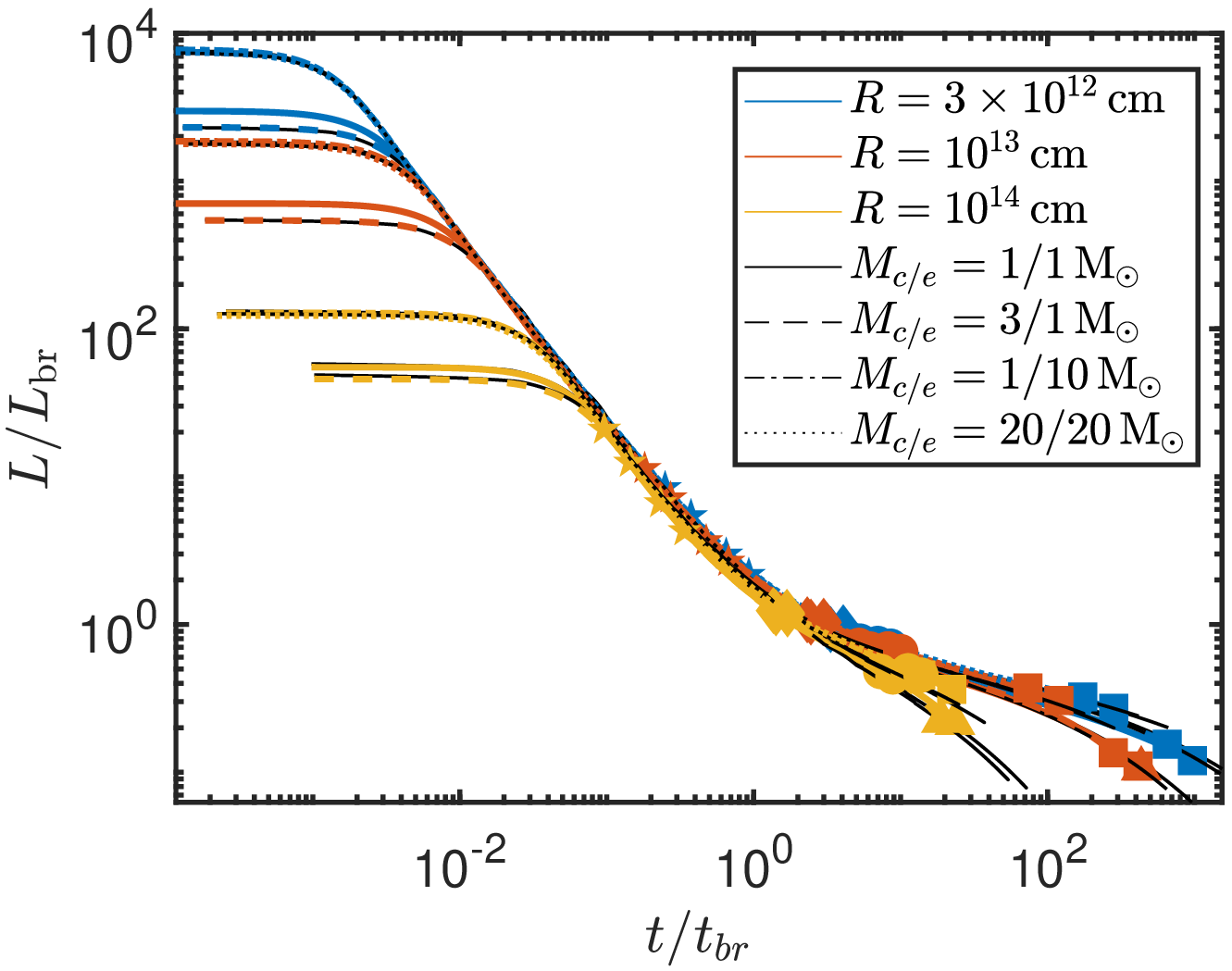}
    \includegraphics[width = \columnwidth]{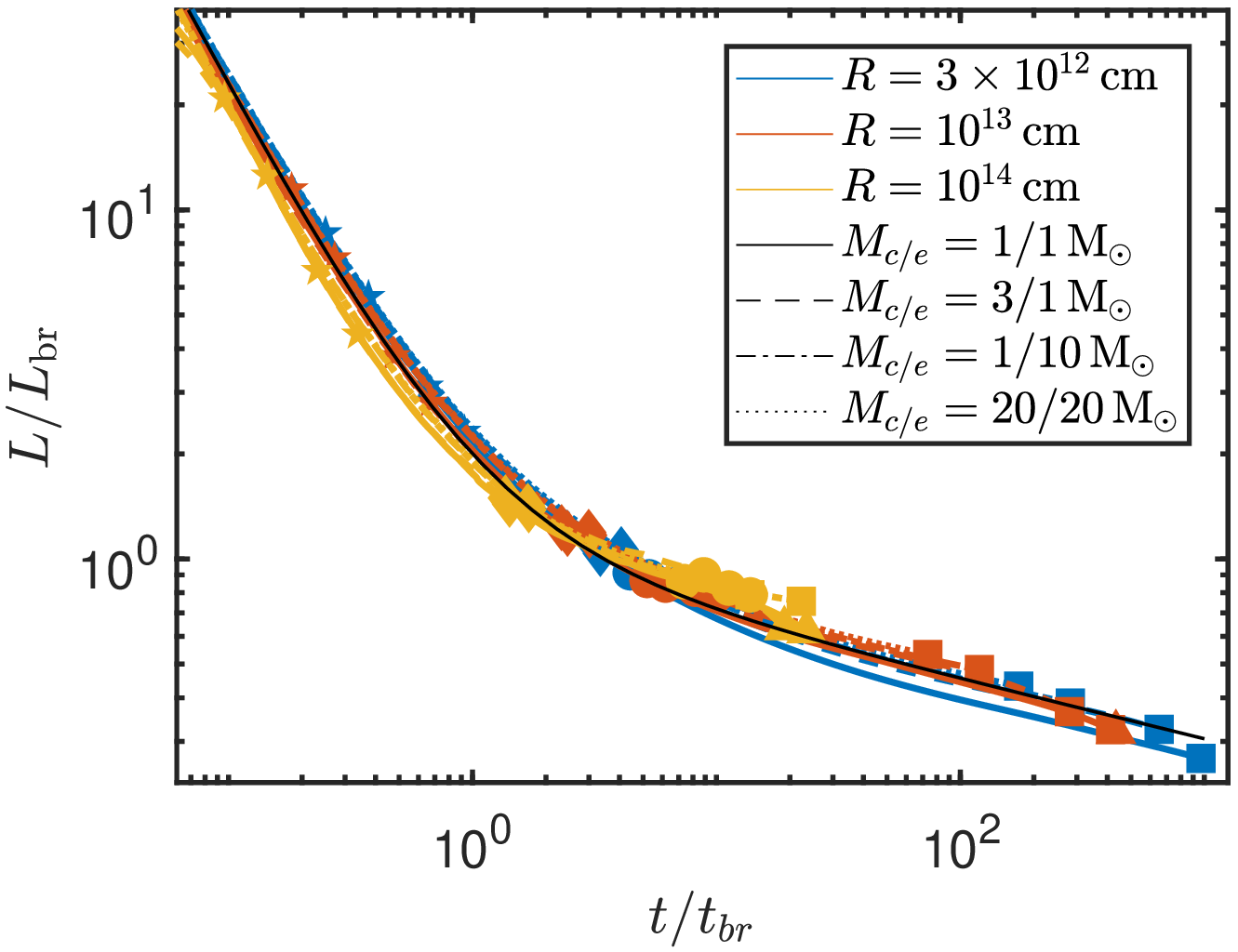}

    \caption{Numerically derived bolometric light curves, for a range of progenitor parameters and explosion energy of $E=10^{51}$~erg, normalized to $\{t_{\rm br}, \,L_{\rm br}\}$ given by Eq.~(\ref{eq:trans_def}). Top panel: The black curve shows the analytic result of Eq.~(\ref{eq:L_sum}) or~(\ref{eq:L_trans}), replacing the $\tilde{t}^{-4/3}$ planar phase term with the exact semi-analytic planar solution of \citetalias{sapir_non-relativistic_2011}, which reduces to $\tilde{t}^{-4/3}$ at $t \gtrsim R/c$. Bottom panel: Same as top, multiplying $L$ by the analytic suppression factor of Eq.~(\ref{eq:L_SW17}) to show explicitly the $\tilde{t}^{-0.17}$ late time behavior. Time symbols indicate the same validity times as in Fig. \ref{fig:L_MSW}. Excellent agreement is obtained during both the breakout and shock cooling phases, demonstrating both the validity of our numeric code results and of the universal form of $L(t)$ at $t>3R/c$.}
    \label{fig:L_MSW_over_t_transition}
\end{figure}

Figures~\ref{fig:L_MSW}-\ref{fig:L_MSW_over_t_transition} present a comparison of the analytic $L(t)$ with the numerically derived $L(t)$ for a wide range of progenitor parameters and explosion energies. In Fig.~\ref{fig:L_MSW} we include the effect of light-travel time, finding that it affects the luminosity by $<10\%$ at $t=3R/c$, and by a smaller amount at later time. Figure~\ref{fig:L_MSW_over_t_transition} shows that normalizing $L$ and $t$ by $L_{\rm br}$ and $t_{\rm br}$ brings all light curves into good agreement. The fractional deviations of the numeric results from the analytic description are typically $<15\%$, with time averaged rms of $\approx 10\%$ over the validity time defined above. Solar metalicity was used in all calculations shown in figures~\ref{fig:L_MSW}-\ref{fig:L_MSW_over_t_transition}. We also observe negligible dependence of $L$ on core structure, with exception of the largest stellar radii $R=10^{14}$ cm, where there is up to a 10\% effect when the core radius is increased by 2 orders of magnitude.

\begin{figure}
    \includegraphics[width = \columnwidth]{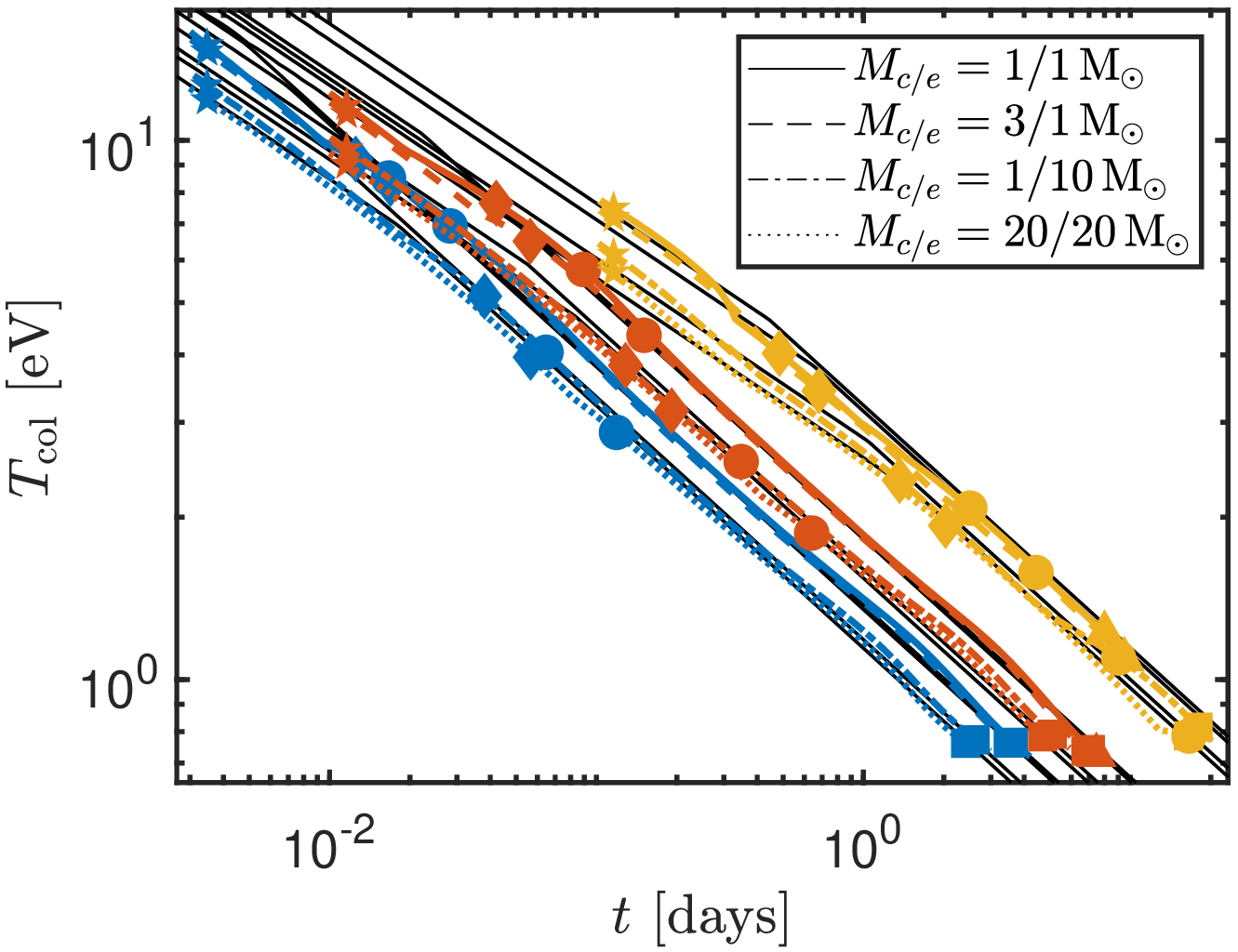}
    \includegraphics[width = \columnwidth]{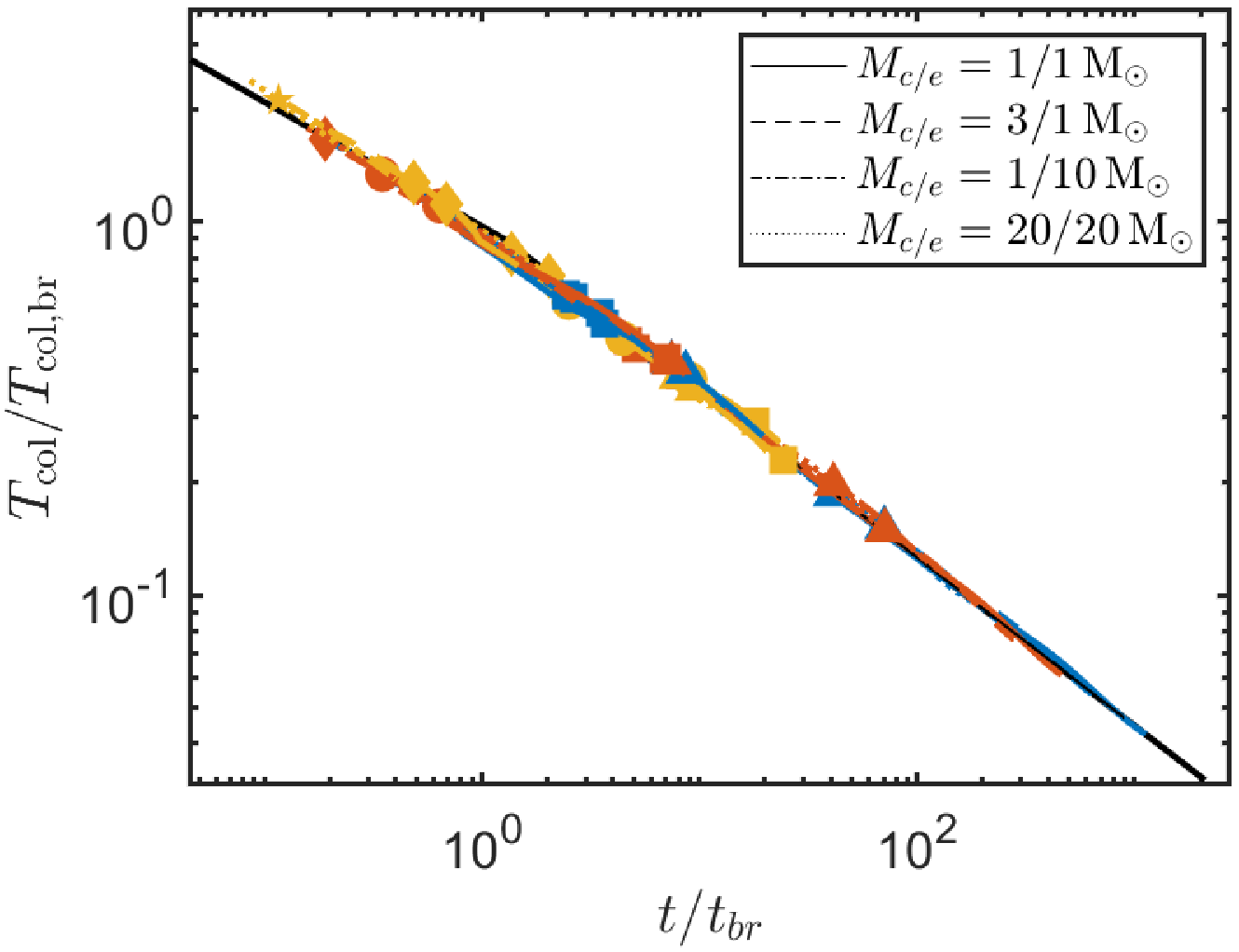}
    \caption{Top panel: Numerically derived color temperatures shown in color for different progenitor radii and mass combinations, explosion energy of $E=10^{51} \rm \, erg$, and solar metallicity, compared to the analytic results, Eq.~(\ref{eq:T_trans}), in black. Bottom panel: Normalizing $T_{\rm col}$ and $t$ to $T_{\rm br}$ and $t_{\rm br}$. Line colors, styles, and validity time marks are the same as in Fig.~\ref{fig:L_MSW}.}
    \label{fig:T_MSW_over_days_and_t_br}
\end{figure}

\begin{figure}
    \includegraphics[width = \columnwidth]{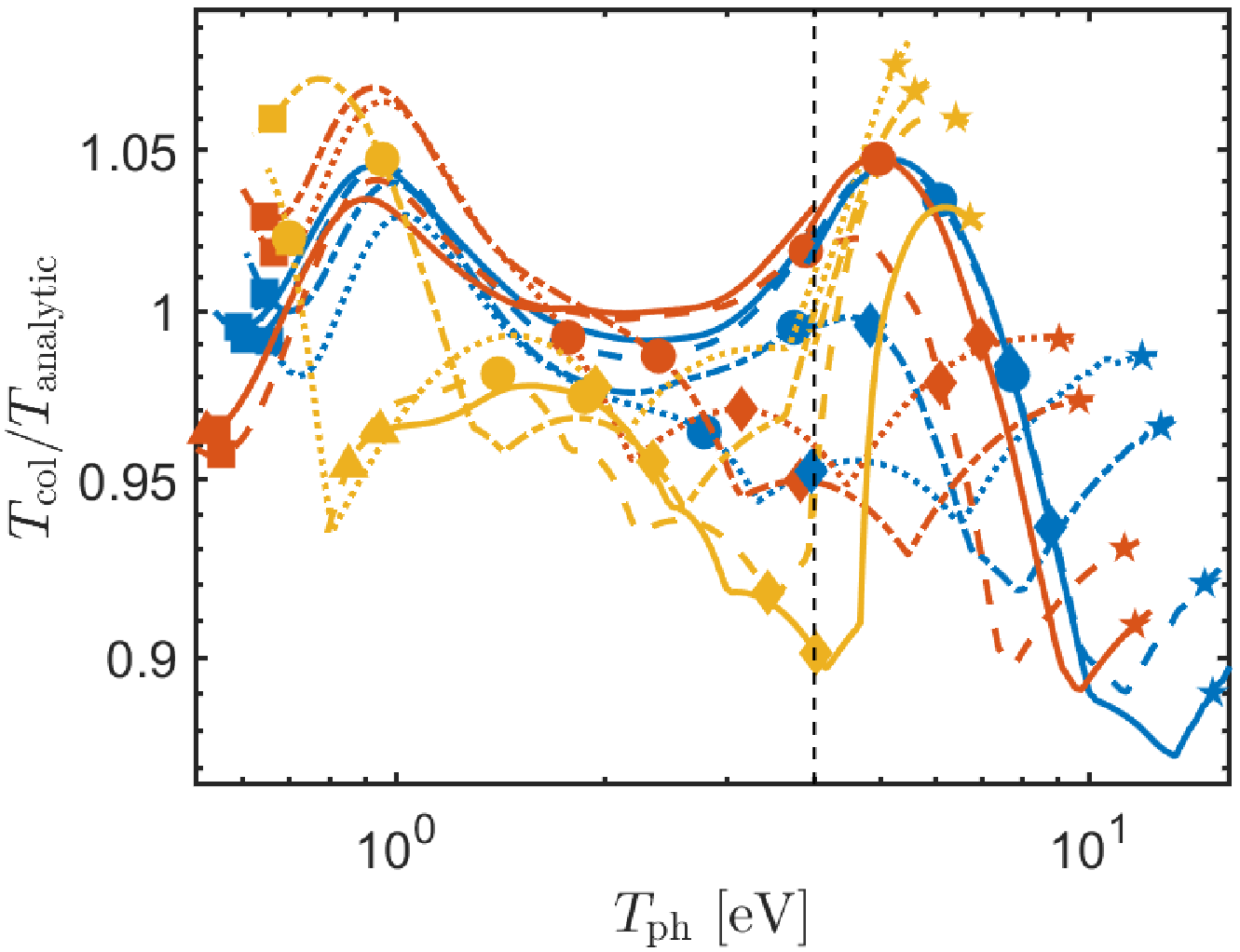}
    \includegraphics[width = \columnwidth]{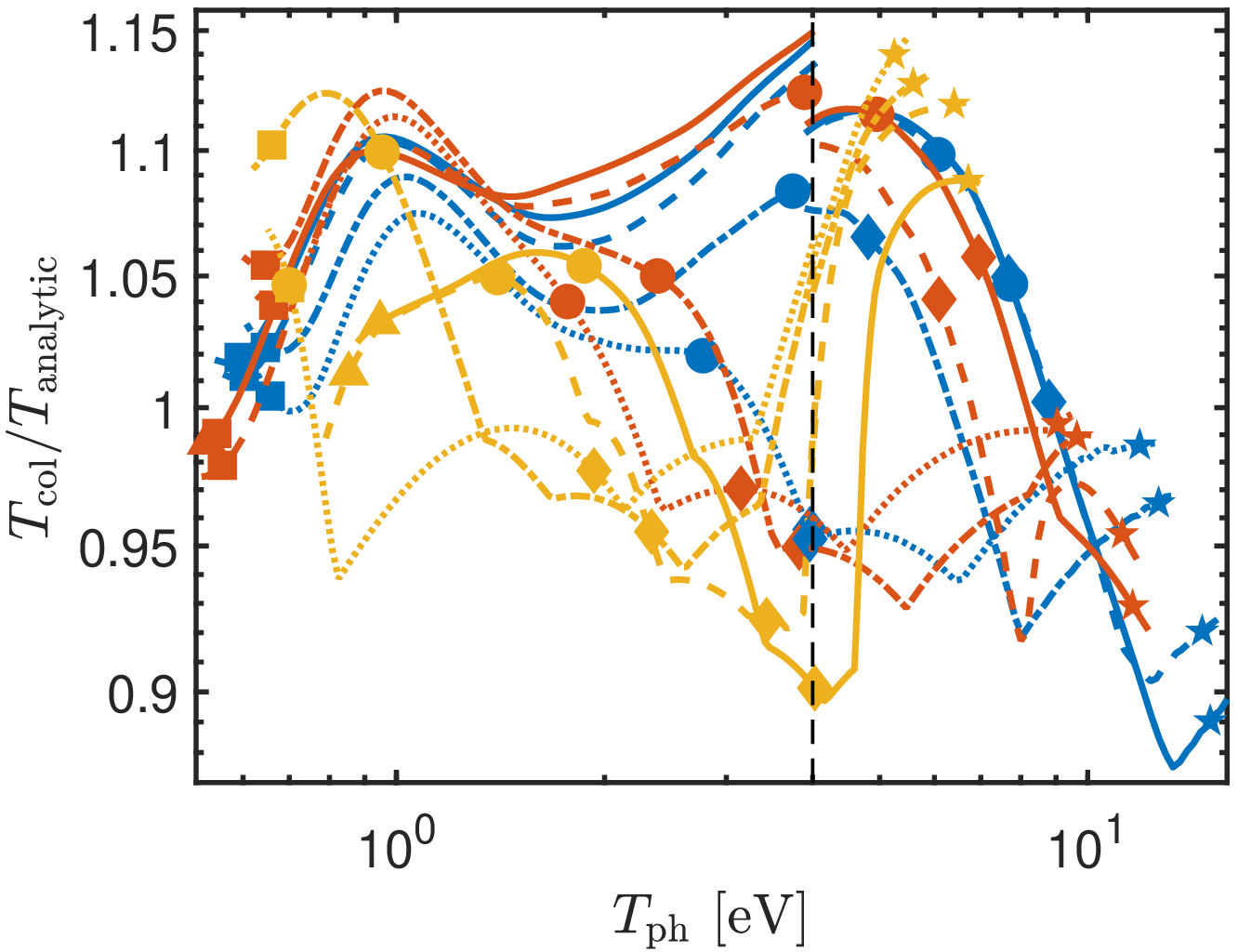}

    \caption{The ratio of numerically derived color temperatures and our analytic result, Eq.~(\ref{eq:T_trans}), for different progenitor radii and mass combinations ($M_{c/e}$ denote core and envelope masses) and explosion energy of $E=10^{51} \rm \, erg$, plotted as a function of the photospheric temperature given by Eq.~(\ref{eq:TphRW}). Top panel: $Z = Z_\odot$, Bottom panel: $Z = 0.1 Z_\odot$. Line colors and styles and validity time marks are the same as in Fig.~\ref{fig:L_MSW}. The vertical dashed line denotes the 4~eV transition between our opacity table and TOPS (see text).}
    \label{fig:T_MSW_over_T_analytic_4eV_w_metallicity}
\end{figure}

\begin{figure}
    \centering
    \includegraphics[width = \columnwidth]{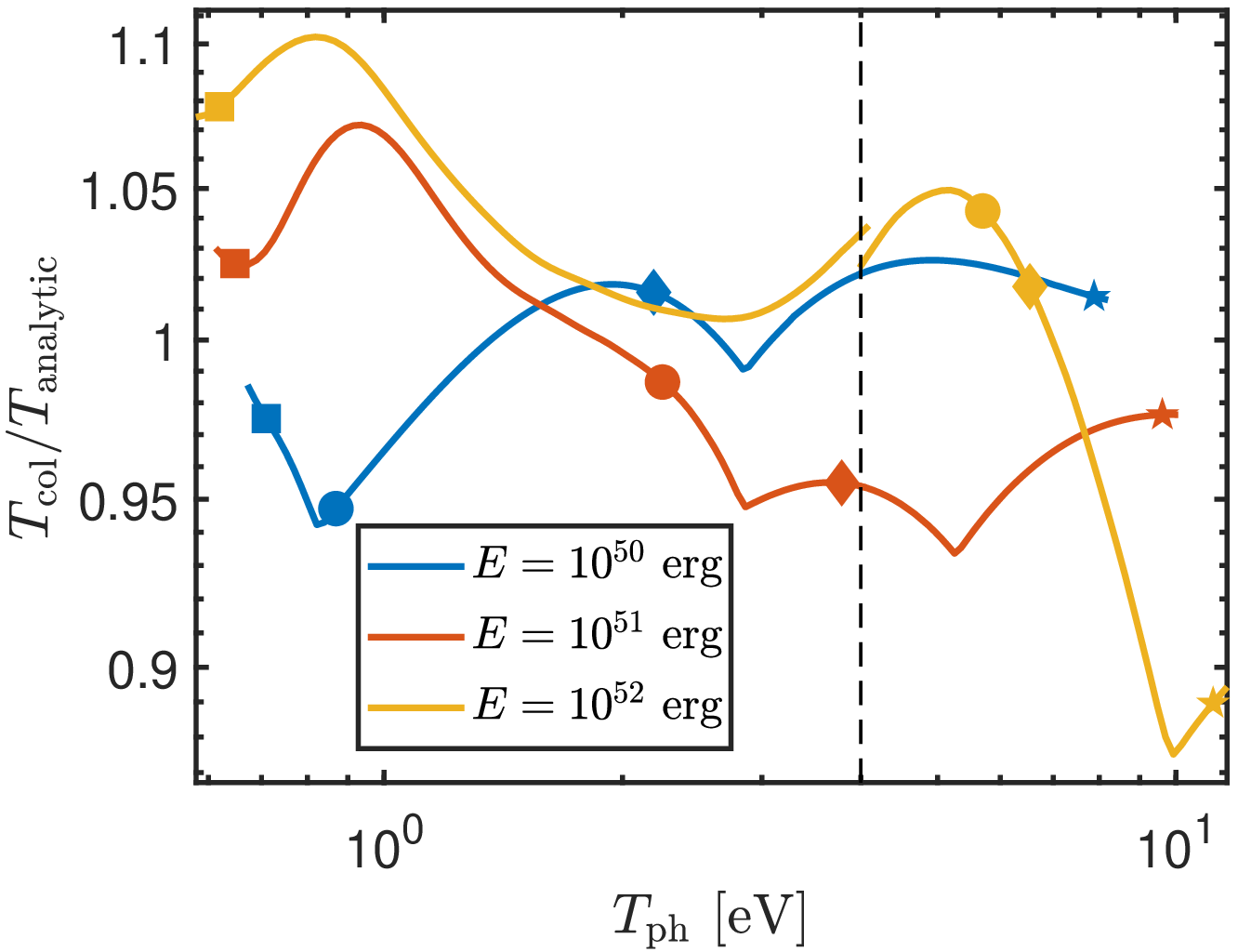}    
    
    \caption{Same as Fig. \ref{fig:T_MSW_over_T_analytic_4eV_w_metallicity} with solar metalicity, for varying explosion energies and $R=10^{13} \rm cm$, $M_{\rm c}=M_{\rm env}=10M_\odot$. Validity time marks are the same as in Fig.~\ref{fig:L_MSW}. The vertical dashed line denotes the 4 eV cutoff between our opacity table and TOPS.}
    \label{fig:Eser Tcol}
\end{figure}

Figures \ref{fig:T_MSW_over_days_and_t_br}-\ref{fig:Eser Tcol} present a comparison of the analytically derived $T_{\rm col}$ with the results of numerical calculations spanning a wide range of progenitor radii, masses, explosion energies, and metallicities. Normalizing $T_{\rm col}$ and $t$ by $T_{\rm br}$ and $t_{\rm br}$ brings all temperature curves into good agreement over the relevant time range. The fractional deviations of the numeric results from the analytic description over all simulations are $\lesssim10\%$, with time averaged rms in all simulations of $\approx 5\%$. As shown in Fig.~\ref{fig:T_MSW_over_T_analytic_4eV_w_metallicity}, the results are insensitive to the metallicity in the range $Z=0.1-1Z_\odot$. This is due to the fact that at the relevant temperatures and densities line opacity is dominated by H and He. We also find similar results for $L$ and $T_{\rm col}$ for slightly lower mass envelopes ($M_{\rm env}/M_{\rm c}=0.1$ - not shown in the figures).

\subsection{Comparison to earlier works}
\label{sec: Previous works}
\begin{figure}
    \includegraphics[width =\columnwidth]{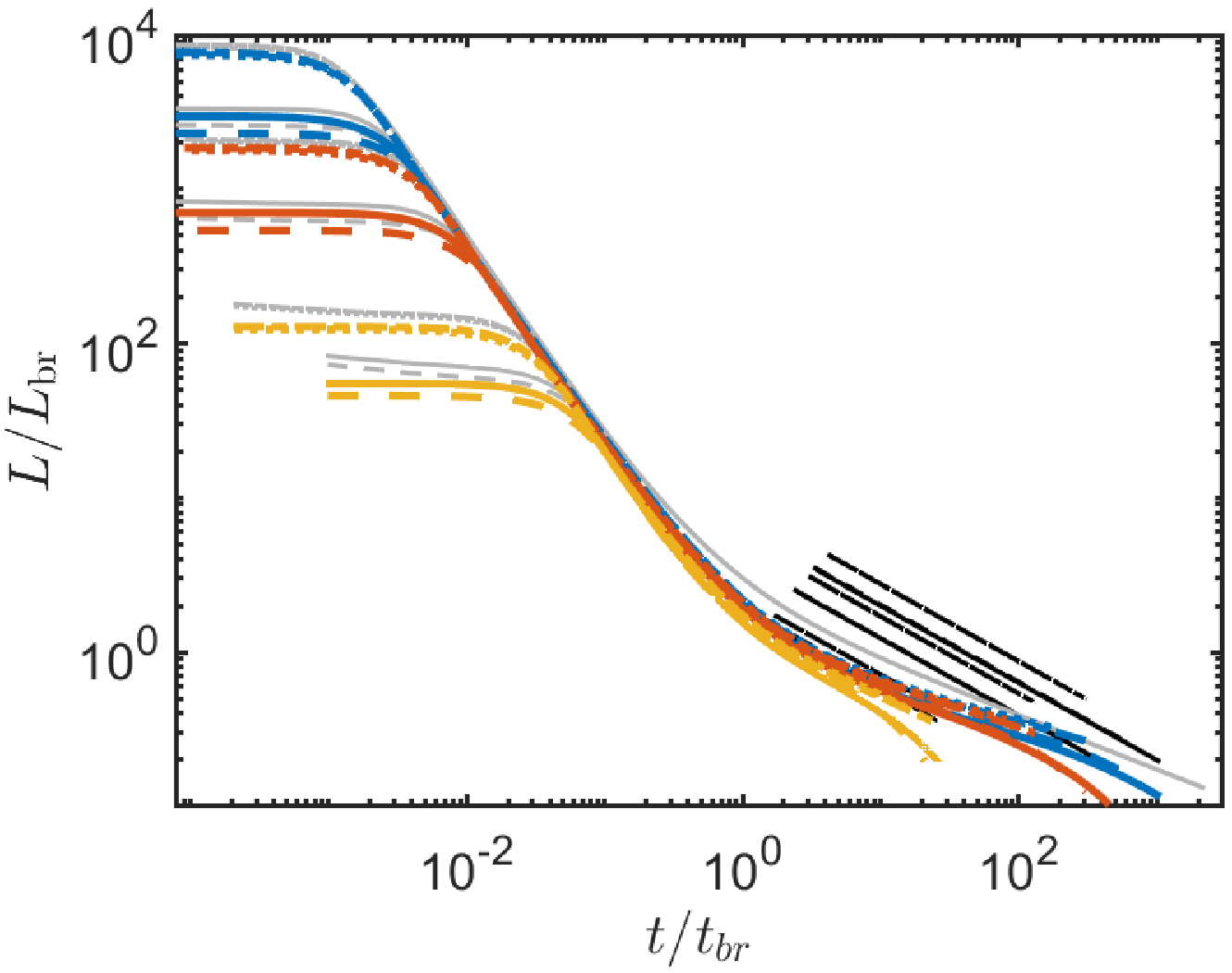}
    \includegraphics[width = \columnwidth]{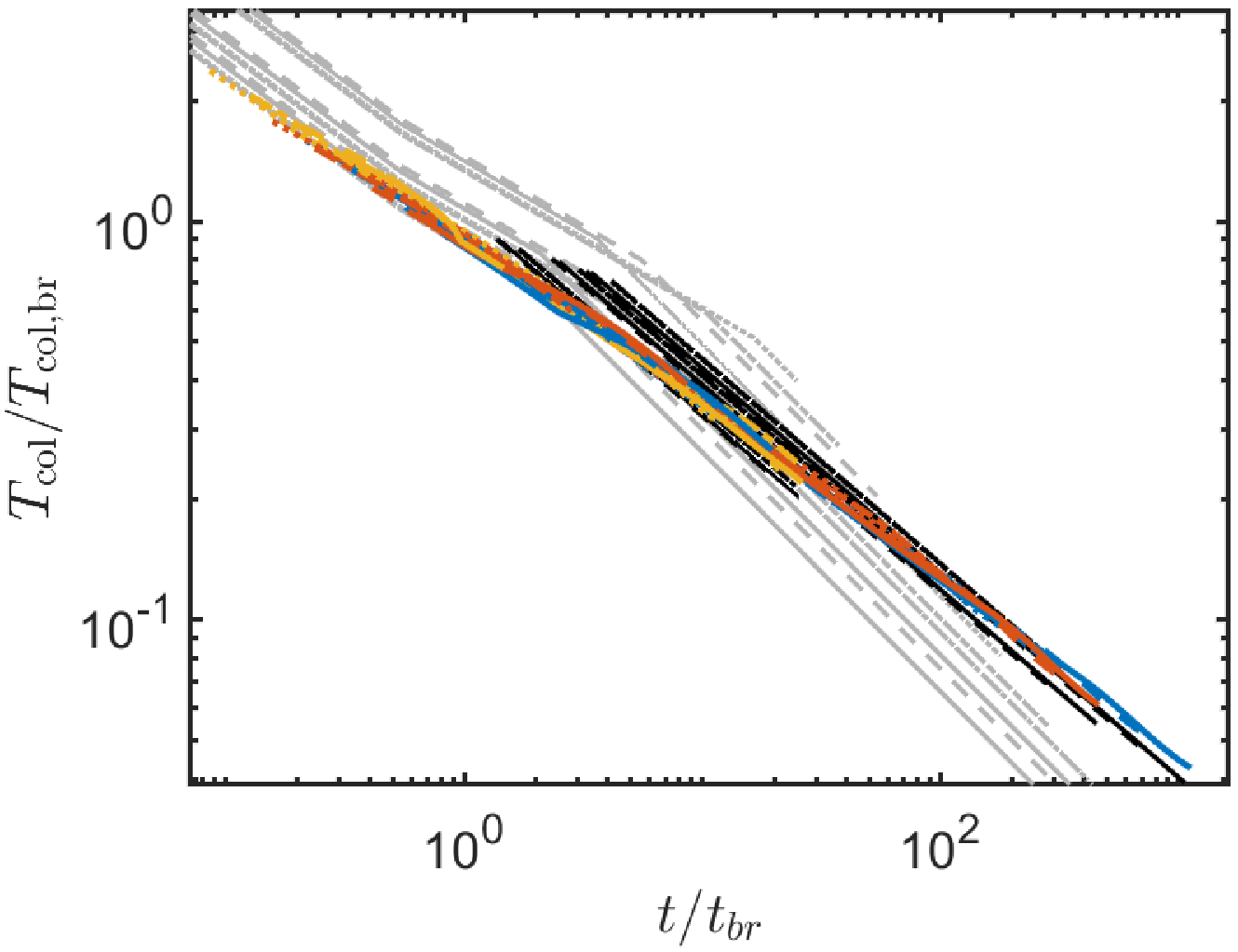}
    \caption{A comparison of the analytic results of \citetalias{shussman_type_2016} (in gray) and \citetalias{piro_shock_2021} (in black) with the results of our numeric calculations (in color - line styles and colors are the same as in Fig. \ref{fig:L_MSW}). The results of \citetalias{piro_shock_2021} for $T_{\rm col}$ are accurate to $\sim25\%$, while they overestimate $L$ by a factor of a few. \citetalias{shussman_type_2016} overestimate $L$ by a factor of $\approx 2$ during the transition to the spherical phase, and their $T_{\rm col}$ deviates from our results by $\approx40\%$.}
    \label{fig: SVN BB compare}
\end{figure}

Our results are compared in Fig.~\ref{fig: SVN BB compare} to the analytic approximations for shock cooling emission derived by \citetalias{shussman_type_2016} for the planar and spherical phases, and with that derived by \citetalias{piro_shock_2021} for the spherical phase.

\citetalias{piro_shock_2021} consider a homologous expansion of plasma with a broken power-law density profile, which approximates the density profile expected for shock driven expansion of a polytropic envelope, and approximate $T_{\rm col}$ by the photosphere temperature (the plasma temperature at Thomson scattering optical depth $\tau_T = 2/3$). Their analytic expressions are given in terms of $R$, the mass of the "extended material" (our "envelope") $M_{\rm env}$, and the energy of the extended material, $E_{\rm env}$. When comparing our numeric calculation results to their analytic results, we extract $E_{\rm env}$ from our numeric calculations. We find that the results of \citetalias{piro_shock_2021} for $T_{\rm col}$ are accurate to $\sim25\%$, while they overestimate $L$ by a factor of a few.

For the comparison with the results of \citetalias{shussman_type_2016}, we extract the values of $\rho_{\rm bo}$, $v_{\rm bo}$ for each of our numeric calculations, and compare the results of the numeric calculation with their analytic formulas, that give $L$ and $T_{\rm col}$ in terms of $\rho_{\rm bo}$, $v_{\rm bo}$ and $R$. We find that they overestimate $L$ by a factor of $\approx 2$ during the transition to the spherical phase \citep[consistent with the results of][]{kozyreva_shock_2020}, and that their $T_{\rm col}$ deviates from our results by $\approx40\%$. As mentioned in the introduction,  \citetalias{shussman_type_2016} also find that the specific flux exceeds in Raleigh-Jeans regime the flux given by a blackbody at $T=T_{\rm col}$ by an order of magnitude, while we find (see \S~\ref{sec:gray_MG_compare}) that the flux is close to blackbody in this regime \citep[consistent with the results of][]{kozyreva_shock_2020}.

\section{Comparison of the analytic model with the results of multi-group calculations}
\label{sec:gray_MG_compare}

We assess the validity of the black-body, at color-temperature $T_{\rm col}$, description of the spectrum by comparing our analytic model approximation with the results of radiation hydrodynamics calculations based on \citet{sapir_numeric_2014}, approximating photon transport by multi-group photon diffusion. The continuous energy distribution of the photons is approximated by a set of photon energy bins. The emission, absorption and scattering of the photons in each energy bin is treated independently with energy dependent cross sections, which are appropriately averaged over the energy in each bin. This enables us to account for deviations from a black-body photon distribution. The numeric code and the results of the multi-group calculations are discussed in detailed in Paper II. Here we give only a brief comparison with the analytic results of this paper.

\begin{figure}
    \centering
    \includegraphics[width = \columnwidth]{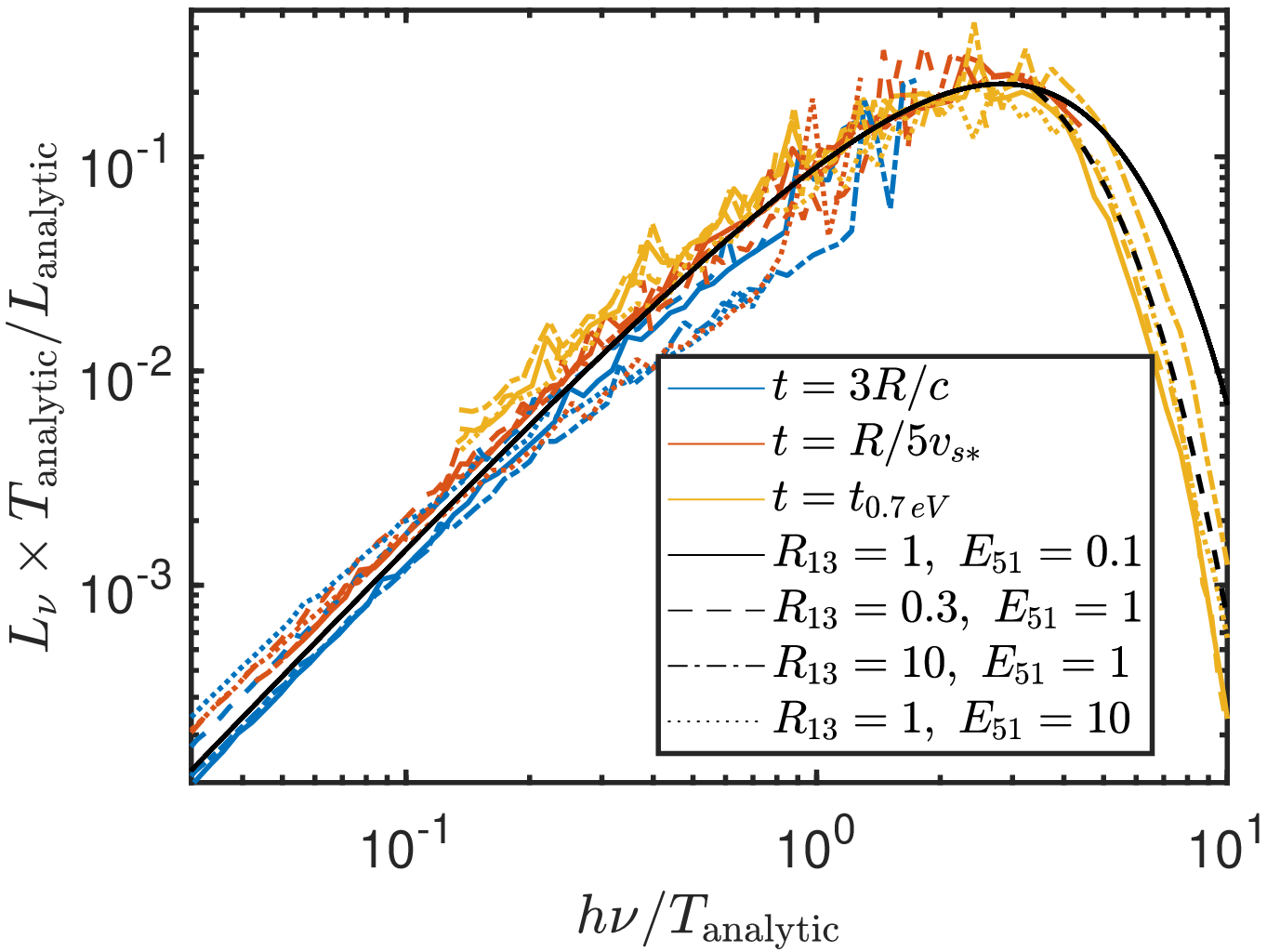}
    
    \caption{A comparison of the specific luminosity, $L_\nu=dL/d\nu$, obtained in multi-group calculations (in color) for progenitor masses $M_{\rm env} = M_{\rm c}=10M_\odot$ and a range of radii and explosion energies. These are compared with our analytic gray formula, Eqs.~(\ref{eq:L_trans}),~(\ref{eq:T_trans}) and (\ref{eq:L_nu_BB_formula}) (in black), and normalized using analytic $L(t)$ and $T_{\rm col}(t)$, in such a way that the gray formula remains constant in the plot at all times (see axis labels). Since $L_\nu(t)$ is in good agreement with our gray formula, this has the effect of collapsing all numeric results onto roughly one curve, without additional fitting required. Different colors correspond to different times, while different line styles correspond to different radii and explosion energies. $L_\nu$ is close to a $T=T_{\rm col}$ Planck spectrum at frequencies below the Planck peak, and suppressed at higher frequencies due to line opacity. The suppressed spectrum is well approximated by a blackbody with $T=0.74 T_{\rm col}$ (black dashed line), as given by Eq.~(\ref{eq:L_nu_BB_mod_formula}). The figure shows the numerical $L_\nu$ only up to  $h\nu=10$~eV, reflecting the range of observable frequencies.}
    \label{fig: Lx_REser}
\end{figure}

In Fig.~\ref{fig: Lx_REser} we show the ratios between the specific luminosities, $L_\nu=dL/d\nu$, obtained in our multi-group calculations and those given by the analytic expressions of \S~\ref{sec:MSW formula} for a sample of progenitor parameters and explosion energy values. The analytic model provides a good approximation to the multi-group results, with deviations limited to tens of percent at wavelengths longer than the peak wavelength. At shorter wavelength, the large opacity due to bound-bound and bound-free transitions lead to a significant suppression of the luminosity. The UV line suppression may be approximately described by modifying the model black-body spectrum (eq.~\ref{eq:L_nu_BB_formula}) according to (see details in \citetalias{morag_shock_2022-1})
\begin{equation}
        L_\nu = L \times \min \left[ \, \frac{\pi B_\nu(T_{\rm col})} {\sigma T_{\rm col}^4} \, , \, \frac{\pi B_\nu(0.74 T_{\rm col})} { \sigma (0.74 T_{\rm col})^4} \right].
 \label{eq:L_nu_BB_mod_formula}
\end{equation}
The $0.74 T_{\rm col}$ term on the RHS suppresses the flux at $h\nu>3T$, and is calibrated to the MG simulations at observable frequencies, $h\nu<10 eV$. Its effect is important at later time, after the Planck peak enters the observational window.

\begin{figure}
    \includegraphics[width = \columnwidth]{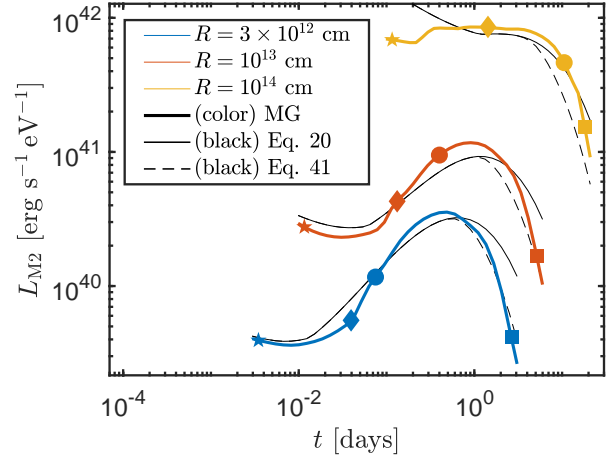}
    \includegraphics[width = \columnwidth]{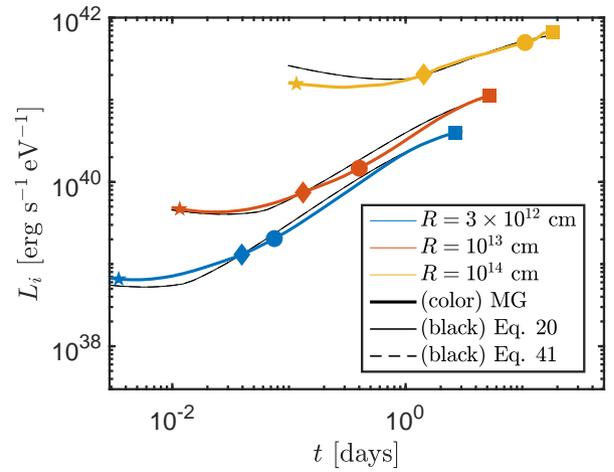}
    \caption{A comparison of light-curves in different bands obtained from the multi-group calculations (solid color lines) with the analytic black-body model of Eqs.~ (\ref{eq:L_trans}),~(\ref{eq:T_trans}) and (\ref{eq:L_nu_BB_formula}) (solid black lines), and with the analytic model including the UV suppression term given by Eq.~(\ref{eq:L_nu_BB_mod_formula}) (dashed black lines). Results are shown for explosions of progenitors with different radii and with $M_{\rm env} = M_{\rm c}=10M_\odot$, $E=10^{51}$ erg, and solar metallicity. Top panel: The Swift/UVOT - UVM2 filter, centered at $0.22\,\mu m$. Bottom panel: SDSS i filter centered at $0.75\,\mu m$.}
    \label{fig:Rser_LC_paper_I}
\end{figure}

\begin{figure}
\includegraphics[width = \columnwidth]{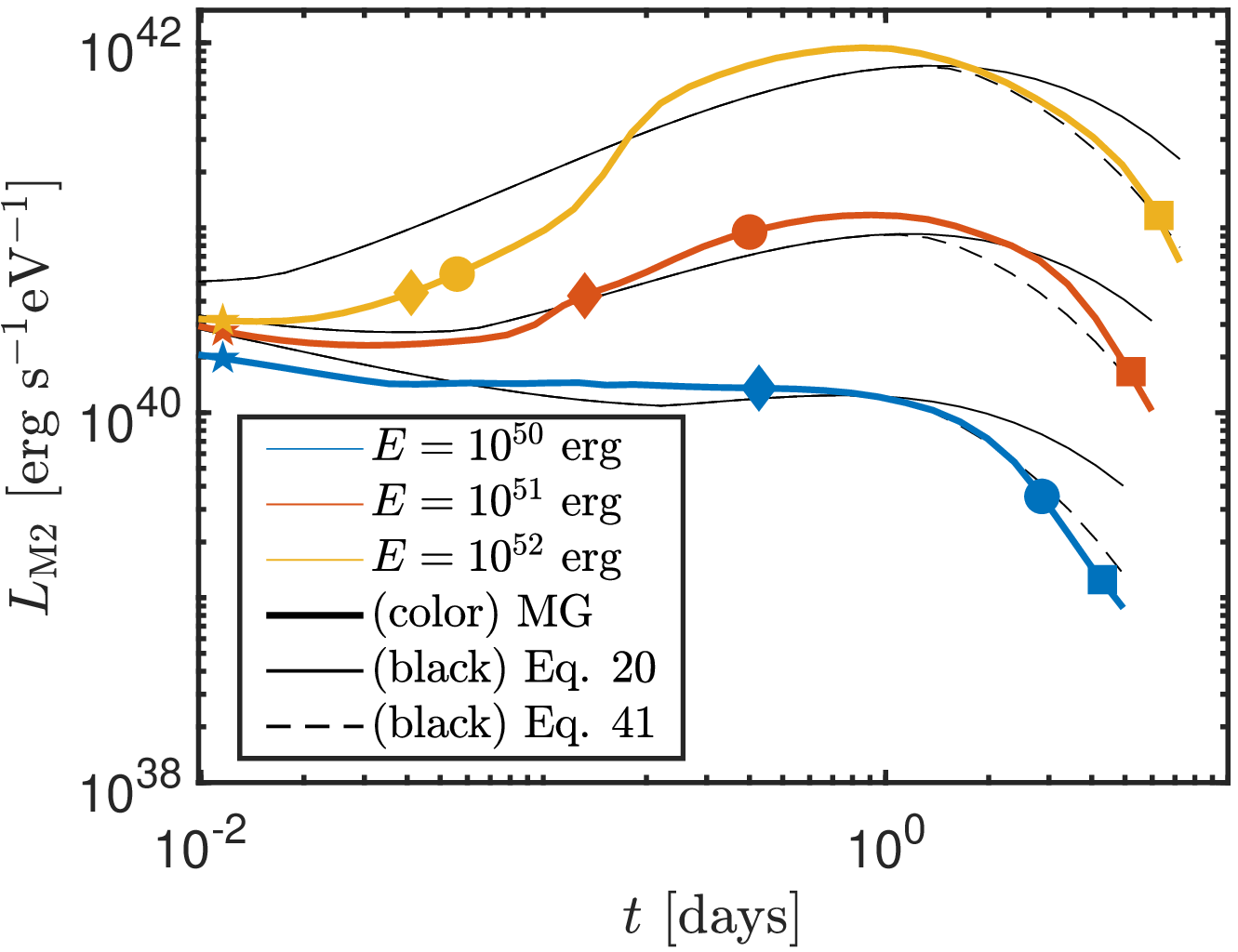}
\includegraphics[width = \columnwidth]{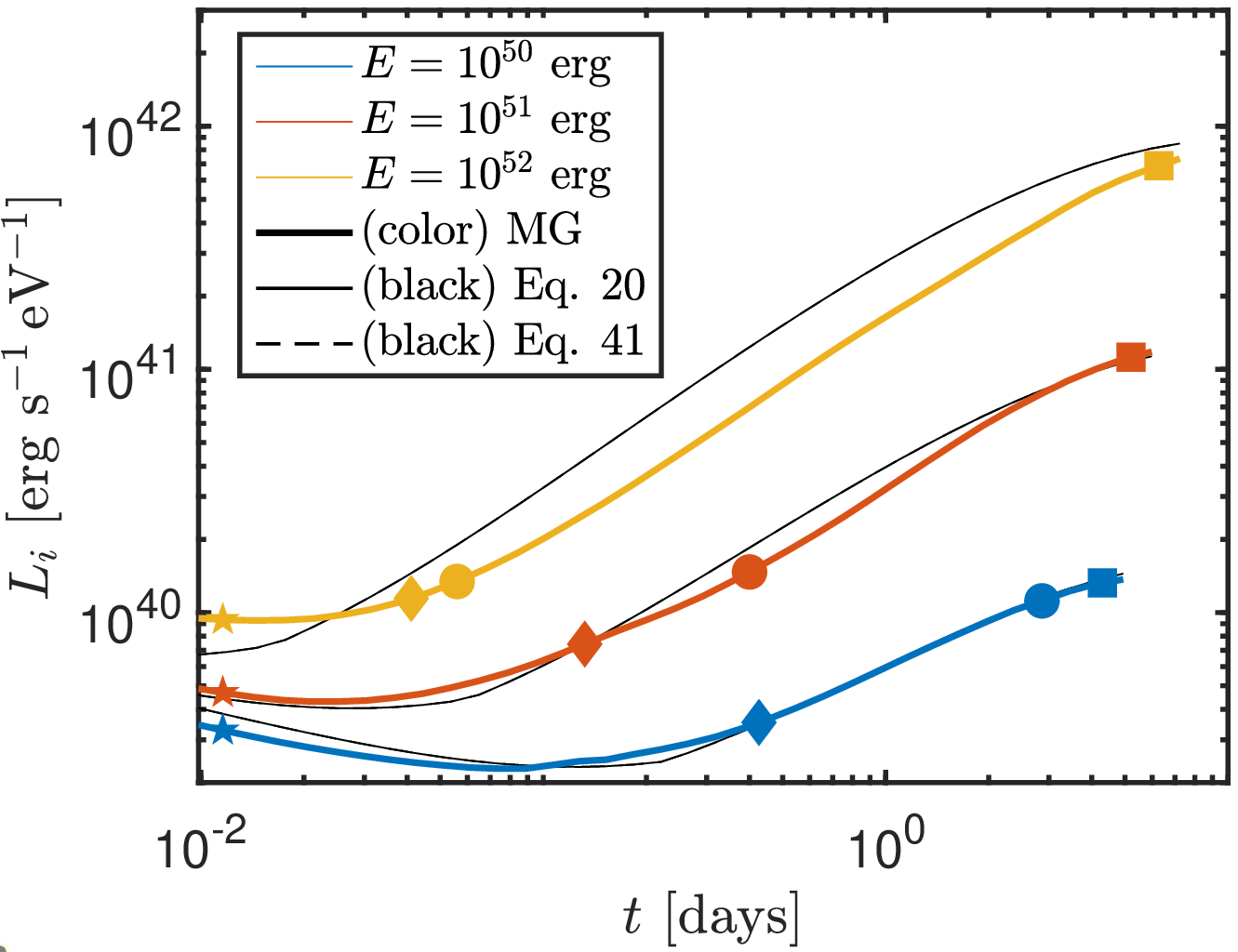}
\caption{Same as Fig. \ref{fig:Rser_LC_paper_I} for explosions of varying energy and progenitors with $R=10^{13} \rm \, cm$,  $M_{\rm env} = M_{\rm c}=10M_\odot$, and solar metalicity.}
\label{fig:Eser_LC_paper_I}
\end{figure}

Figures~\ref{fig:Rser_LC_paper_I} and~\ref{fig:Eser_LC_paper_I} show a comparison of light-curves in specific bands, from the IR to the UV, obtained from the multi-group calculations and from the analytic model. The analytic model provides a good approximation of the multi-group results. The line suppression of the flux at high frequencies leads mainly to a faster decline of the UV flux after the peak in the light-curve.

\section{Discussion and Summary}
\label{sec: Summary}

We have derived a simple analytic description of the bolometric luminosity, $L$, and color temperature, $T_{\rm col}$, of shock cooling emission produced by the explosions of red supergiants (RSG) composed of compact cores surrounded by polytropic envelopes. The analytic
is based on an interpolation between the analytic results of \citetalias{sapir_non-relativistic_2011} and \citetalias{katz_non-relativistic_2012} for the planar phase of the expansion, and of \citetalias{rabinak_early_2011}/\citetalias{sapir_uv/optical_2017} for the later spherical phase. It is given by Eqs.~(\ref{eq:L_sum})-(\ref{eq:T_col_MSW_2}) in \S~\ref{sec:MSW formula}, and is valid from post breakout, $t>3R/c$, up to H recombination, $T\approx0.7$~eV, or until the photon diffusion time through the envelope becomes shorter than the dynamical time, see Eq.~(\ref{eq:t_lc1}). A concise summary of all the equations required for constructing $L(t)$ and $T_{\rm col}$, and for determining the validity time of the model, is given in the appendix.

We have presented in \S~\ref{sec:numeric_res} a comparison of the analytic model to the results of numerical calculations of shock cooling emission for a wide range of explosion energies,  $E=10^{50}-10^{52}$~erg, and progenitor parameters: masses in the range $M=2-40M_{\odot}$, radii $R=3\times10^{12}-10^{14}$~cm, stellar core radii and masses $R_{\rm c}/R=10^{-1}-10^{-3}$ and $M_{\rm env}/M_{\rm c}=10-0.1$, and metallicities $Z=0.1Z_{\odot}-Z_{\odot}$. The numerically derived $L$ and $T_{\rm col}$ are described by the analytic expressions with $10\%$ and $5\%$ accuracy respectively over the explored parameter range, as demonstrated in figures~\ref{fig:L_MSW}-\ref{fig:Eser Tcol}. A concise description of the comparison of our results to earlier work appears in figure~\ref{fig: SVN BB compare}.

In the numerical calculations presented in \S~\ref{sec:numeric_res}, photon transport was approximated by diffusion with constant opacity, corresponding to electron scattering of a highly ionized plasma. $T_{\rm col}(t)$ was obtained from the hydrodynamic profiles as the plasma temperature at the "thermal depth", from which photons diffuse out of the envelope without further absorption. The thermal depth was determined by Eq.~(\ref{eq:Itais_Prescription}), using time and space dependent effective "gray" (frequency independent) opacity, based on opacity tables that we have constructed for this purpose, as described in \S~\ref{sec:kappa}. 

In an accompanying paper (Paper II) we show, using a large set of multi-group photon diffusion calculations, that the spectral energy distribution is well described by a Planck spectrum with $T=T_{\rm col}$, except at UV frequencies (beyond the spectral peak at $3T_{\rm col}$), where the flux is significantly suppressed due to the presence of strong line absorption. A brief comparison of the analytic results with the results of multi-group calculations is presented in figure~\ref{fig: Lx_REser}. An analytic description of the specific luminosity, $dL/d\nu$, including the UV suppression is given by Eq.~(\ref{eq:L_nu_BB_mod_formula}). Figures~\ref{fig:Rser_LC_paper_I} and~\ref{fig:Eser_LC_paper_I} show a comparison of light-curves in specific bands, from the IR to the UV, obtained from the multi-group calculations and from the analytic model. The analytic model provides a good approximation of the multi-group results. The line suppression of the flux at high frequencies leads mainly to a faster decline of the UV flux after the peak in the light-curve. A detailed discussion of the deviations from a thermal spectrum will be given in Paper II, including an analytic description of the small ($\sim10\%$) deviations in the Rayleigh-Jeans regime. 

The agreement between the analytic and the multi-group results demonstrates the validity of the approximation $\kappa_{abs}=\kappa_{R}-\kappa_{es}$, used (following \citetalias{rabinak_early_2011}) in Eq.~(\ref{eq:Itais_Prescription}) for the derivation of the color temperature $T_{\rm col}$. It also also confirms the importance of including the bound-free and bound-bound contributions to the opacity- we find a ratio $f_{T}\approx1.1$ between the color and photospheric temperatures, consistent with \citetalias{rabinak_early_2011,sapir_uv/optical_2017}, while $f_{T}\approx2$ is obtained if only the free-free contribution to the opacity is considered. 

The analytic model is completely determined, see appendix, by four parameters: $\{R,v_{\rm s\ast},f_\rho M,M_{\rm env}\}$- $R$ is the progenitor radius; $v_{\rm s\ast}$ is the characteristic shock velocity, which is related (and close) to the characteristic ejecta velocity $v_\ast=\sqrt{E/M}$, where $E$ and $M$ are the ejecta mass and energy (see Eq.~(\ref{eq:vstar})); $M_{\rm env}$ is the envelope mass; $f_\rho$ (see Eq.~(\ref{eq:rho_in})) is a numerical factor  describing the envelope structure, of order unity for convective envelopes, $f_\rho\approx(M_{\rm env}/M_{\rm c})^{1/2}$ \citepalias{sapir_uv/optical_2017}. The breakout and shock cooling emission depend at early time, $t\ll t_{\rm tr}$, mainly on the radius $R$ and shock velocity $v_{\rm s\ast}$. These parameters may therefore be well constrained by the observations. On the other hand, the dependence of the emission on $f_\rho M$, and on the progenitor structure in general (see sensitivity to core structure - \S~\ref{sec:numeric_res}, and deviations from polytrope - \S~\ref{current paper}), is very weak. This parameter cannot therefore be inferred reliably from observations, an issue that we address in upcoming works. Finally, the suppression of the luminosity at late time, $t\sim t_{\rm tr}$ (when photons are emitted from deep within the envelope) may enable one to determine $t_{\rm tr}$ and to infer $M_{\rm env}$ the envelope mass, using Eq.~(\ref{eq:t_transp}) and given $v_{\rm s\ast}$.

\section*{Acknowledgements}
We thank Ido Irani and Gilad Sadeh for insightful discussion and the anonymous reviewer for their comments. EW's research is partially supported by ISF, GIF and IMOS grants.

\section*{Data Availability}
Data and numerical codes from this paper will be provided upon reasonable request to the corresponding author. Our opacity table will be released to the public in a subsequent work.


\bibliographystyle{mnras}
\bibliography{references}

\appendix

\section{Summary of model equations}
\label{appendix}

$L$ and $T_{\rm col}$ are well approximated by
\begin{equation}
    \label{eq:L_trans - Appendix}
    L/L_{\rm br}=\tilde{t}^{-4/3}+ A\exp\left[-\left(at/t_{\rm tr}\right)^{\alpha}\right]
    \tilde{t}^{-0.17},
\end{equation}
\begin{equation}
\label{eq:T_trans - Appendix}
    T_{\rm col}/T_{\rm col,br}=\min\left[0.97\,\tilde{t}^{-1/3},\tilde{t}^{-0.45}\right],
\end{equation}
with $\tilde{t}=t/t_{\rm br}$ and $\{A,a,\alpha\}=\{0.9,2,0.5\}$. These approximations hold at
\begin{equation}
\label{eq:Ats}
   3 R / c = 17\,R_{13} \, {\rm min} < t < \min[t_{\rm 0.7 eV}, t_{\rm tr}/a].
\end{equation}
 The deviation of the specific luminosity from a black-body spectrum, due to UV line suppression, may be approximately described by 
\begin{equation}
        L_\nu = L \times \min \left[ \, \frac{\pi B_\nu(T_{\rm col})} {\sigma T_{\rm col}^4} \, , \, \frac{\pi B_\nu(0.74 T_{\rm col})} { \sigma (0.74 T_{\rm col})^4} \right].
 \label{eq:AL_nu_BB_mod_formula}
\end{equation}

The various terms in Eqs.~(\ref{eq:L_trans - Appendix}-\ref{eq:Ats}) are given by
\begin{equation}
\label{eq:t_br_of_vs - Appendix}
    t_{\rm br}= 0.86 \, R_{13}^{1.26} v_{\rm s*,8.5}^{-1.13}
(f_{\rho}M_0\kappa_{0.34})^{-0.13}\,\text{hrs},
\end{equation}
\begin{equation}
\label{eq:L_br_of_vs - Appendix}
L_{\rm br}=3.69\times10^{42} \, R_{13}^{0.78} v_{\rm s*,8.5}^{2.11}
(f_{\rho}M_0)^{0.11} \kappa_{0.34}^{-0.89}\,{\rm erg \, s^{-1}},
\end{equation}
\begin{equation}
\label{eq:T_br_of_vs - Appendix}
T_{\rm col,br}= 8.19 \, R_{13}^{-0.32} v_{\rm s*,8.5}^{0.58}
(f_{\rho}M_0)^{0.03} \kappa_{0.34}^{-0.22}\,{\rm eV}.
\end{equation}
\begin{equation} \label{eq:At_0_7eV}
    t_{0.7 \rm eV} = 6.86 \, R_{13}^{0.56} v_{\rm s*,8.5}^{0.16} \kappa_{0.34}^{-0.61} (f_{\rho}M_0)^{-0.06} \rm  \, days
\end{equation}
\begin{equation}
\label{eq:At_transp}
    \begin{split}
        \begin{aligned}
            t_{\rm tr} &= 19.5 \, \sqrt{\frac{\kappa_{0.34}M_{\rm env,0}} {v_{\rm s*,8.5}}}\, \text{days}.
        \end{aligned}
    \end{split}
\end{equation}

Here, $R= 10^{13}R_{13}$~cm, is the progenitor radius, $M=1 M_{0} M_\odot$ is the ejecta mass, $M_{\rm env}= M_{\rm env,0} M_\odot$ is the mass of the envelope, and $v_{\rm s\ast}=10^{8.5}v_{\rm s*,8.5}\,{\rm cm/s}$ is related to the characteristic ejecta velocity by \begin{equation}
\label{eq:Avstar}
    v_{\rm s\ast}\approx 1.05 f_\rho^{-0.19}v_\ast,\quad v_\ast\equiv\sqrt{E/M},
\end{equation}
where $E$ is the energy deposited in the ejecta. $f_\rho$ (see Eq.~(\ref{eq:rho_in}) is a numerical factor, of order unity for convective envelopes, $f_\rho\approx(M_{\rm env}/M_{\rm c})^{1/2}$ \citepalias{sapir_uv/optical_2017}.

The opacity and time are normalized as $\kappa=0.34 \kappa_{0.34} \rm cm^2 g^{-1}$,  
$t=1 \, t_{\rm hr} \, {\rm hour}=1 t_d$~d and we define $t=0$ as the time at which the breakout flux peaks.

\bsp	
\label{lastpage}
\end{document}